\pdfoutput=1

\documentclass[12pt,nohyper,twosided]{JHEP3}

\usepackage{epsfig}
\usepackage{amssymb}
\usepackage{amsmath}
\usepackage{multirow}
\usepackage{mathrsfs}
\usepackage{color}
\usepackage{array}

\usepackage{graphicx}
\usepackage{cite}

\renewcommand{\hat}{\widehat}
\renewcommand{\tilde}{\widetilde}
\definecolor{paper_blue}{rgb}{0.3,0.2,0.75}
\definecolor{paper_red}{rgb}{0.65,0.1,0.15}
\definecolor{paper_green}{rgb}{0.05,0.35,0.125}
\definecolor{paper_grey}{gray}{0.375}

\newcommand{\eq}[1]{\begin{equation}#1\end{equation}}
\newcommand{\eqs}[1]{\begin{equation}\begin{split}#1\end{split}\end{equation}}
\newcommand{\ab}[1]{\langle #1\rangle}
\newcommand{\smallminus}{{\rm\rule[2.4pt]{6pt}{0.65pt}}}
\newcommand{\smallplus}{\hspace{0.5pt}\text{{\small+}}\hspace{-0.5pt}}
\renewcommand{\cap}{\mathrm{\raisebox{0.75pt}{{$\,\bigcap\,$}}}}

\newcommand{\qab}{\text{Q}}
\newcommand{\qabinverse}{\left(\text{Q$_F^{-1}$}\right)}
\newcommand{\egFourBracket}{\mbox{$\ab{\,\cdot\,\,\cdot\,\,\cdot\,\,\,\cdot\,}$}}
\newcommand{\egTwoBracket}{\mbox{$\ab{\,\cdot\,\,\cdot\,}$}}
\newcommand{\mathematica}[3]{\vspace{0.35cm}\noindent\boxed{\begin{minipage}{#1\textwidth}\begin{tabular}{lp{11cm}}
{\color{paper_blue}{\scriptsize{\tt In[1]:=}}}&{\tt #2}\\
{\color{paper_blue}{\scriptsize {\tt Out[1]:=}}}&{\tt #3}
\end{tabular}\end{minipage}}\vspace{0.35cm}}

\title{{\LARGE Efficient Tree-Amplitudes in $\mathcal{N}=4$:\\\Large Automatic BCFW Recursion in {\sc Mathematica}}}
\author{Jacob L. Bourjaily\\Department of Physics, Princeton University, Princeton, NJ 08544, and\\School of Natural Sciences, Institute for Advanced Study, Princeton, NJ 08540}
\preprint{November 2010}

\abstract{We describe an efficient implementation of the BCFW recursion relations for tree-amplitudes in $\mathcal{N}=4$ super Yang-Mills, which can generate analytic formulae for general N$^k$MHV colour-ordered helicity-amplitudes---which, in particular, includes all those of non-supersymmetric Yang-Mills. This note accompanies the public release of the \mbox{{\sc Mathematica}} package {\tt bcfw}, which can quickly (and automatically) generate these amplitudes in a form that should be easy to export to any computational framework of interest, or which can be evaluated directly within {\sc Mathematica} given external states specified by four-momenta, spinor-helicity variables or momentum-twistors. Moreover, {\tt bcfw} is able to solve the BCFW recursion relations using any one of a three-parameter family of recursive `schemes,' leading to an extremely wide variety of distinct analytic representations of any particular amplitude. This flexibility is made possible by {\tt bcfw}'s use of the momentum-twistor Grassmannian integral to describe all tree amplitudes; and this flexibility is accompanied by a remarkable increase in efficiency, leading to formulae that can be evaluated much faster---often by several orders of magnitude---than those previously derived using BCFW.}

\begin{document}

\tableofcontents

\newpage

\section{Introduction}\label{introduction}

The on-shell recursion relations for scattering amplitudes described by Britto, Cachazo, Feng and Witten (BCFW) \cite{Britto:2004ap,Britto:2005fq} are very well known and have been widely used to compute scattering amplitudes for both purely-theoretical and extremely practical purposes in a wide variety of theories \cite{Cohen:2010mi}. They represent one of the major new tools in the study of quantum field theory. Theoretically, the power and simplicity of the recursive definitions of scattering amplitudes has allowed for the development of an arguably `phenomenological' approach to the advancement of our understanding of quantum field theory: by making once intractable problems essentially effortless, many new questions can be asked---and answered. And practically, tree-amplitudes for processes involving many external particles are of importance for the accurate prediction of backgrounds for new physics at the LHC, for example; BCFW---along with a variety of other computational frameworks such as those based on the powerful Berends-Giele recursion relations \cite{Berends:1987me}---has greatly aided this effort. Considering for example that colour-stripped tree-amplitudes in $\mathcal{N}=4$ encode all the data of scattering amplitudes in ordinary, non-supersymmetric massless QCD  \cite{Dixon:2010ik}, it is clear that understanding $\mathcal{N}=4$ is an important step along the way to understanding QFT in general, and as it is observed in the Standard Model as backgrounds for new physics at the LHC.

Partly because of the existence and incredible simplicity of recursive definitions of the S-Matrix, tree-amplitudes in $\mathcal{N}=4$ have been largely understood in the literature for some time now. Indeed, there exists today a large number of independent presentations of all perturbative tree-amplitudes in $\mathcal{N}=4$, including those based on the BCFW recursion relations \cite{ArkaniHamed:2008gz,ArkaniHamed:2010kv,Drummond:2008cr}, twistor string theory \cite{Dolan:2007vv,SV,Bourjaily:2010kw,Skinner:2010cz,Witten:2003nn}, contour integrals in the Grassmannian \cite{ArkaniHamed:2009dn,Kaplan:2009mh}, and the CSW recursion relations \cite{Cachazo:2004kj,ArkaniHamed:2009sx,Bullimore:2010dz}, for example. Many of these results were made possible in part through the existence of privately-developed, powerful computational tools which have proven themselves essential for gaining intuition and necessary for checking results. Recently, some of these tools have become publicly available through the release of the {\sc Mathematica} package {\tt Gluon-Gluino-Trees} ({\tt GGT}), \cite{Dixon:2010ik}, which is capable of analytically computing all N$^k$MHV tree-amplitudes involving combinations of external gluons and gluinos, and can compute these numerically using the package `{\tt S@M},' \cite{Maitre:2007jq}.

With this note, we extend the reach of these resources to include {\it all} N$^k$MHV tree-amplitudes---including those involving squarks---by making available the {\sc Mathematica} package {\tt bcfw}, included with the submission of this paper on the {\tt arXiv}. In addition to its complete generality, there are two principle features of {\tt bcfw} that should make it particularly useful to those who are interested in gaining intuition about or evaluating tree-amplitudes in $\mathcal{N}=4$. First, the analytic formulae generated by {\tt bcfw} are often dramatically more compact and easier to evaluate than any existing formulae obtained using BCFW. To highlight the magnitude of this improvement, \mbox{Table \ref{unpolarized_times}} lists the times required to evaluate unpolarized $n$-gluon scattering cross-sections using {\tt bcfw} and {\tt GGT/S@M}.\footnote{The times quoted in Table \ref{unpolarized_times} (and elsewhere in this note) were obtained using the author's Apple laptop computer, which has a 2.6 GHz Intel Core 2 Duo processor.} As will be discussed at greater length in \mbox{section \ref{comparison_section}}, these gains in efficiency can be traced directly to {\tt bcfw}'s: 1.\ \nolinebreak use of momentum-twistor variables, and 2.\ representation of all tree-amplitudes in a fully-supersymmetric way (realized as contour integrals over the Grassmannian), making any $n$-point N$^k$MHV helicity-amplitude easily obtained from any other. Another feature of {\tt bcfw} that should make it useful to researchers is its ability to solve the BCFW-recursions using a wide variety of different recursive `schemes,' leading to a large number of independent analytic formulae for any particular amplitude.\footnote{For example, we have included as a worked example in the demonstration file included with the {\tt bcfw} package the construction of all $74$ linearly-independent, 20-term formulae for the \mbox{8-point} N$^2$MHV tree-amplitude, involving a total of 176 different Yangian-invariant objects.} And it may be worth mentioning that the {\tt bcfw} package has been designed with hopes of being intuitive-enough to be useful even to those with very little experience with {\sc Mathematica}.
\begin{table}[h!]\vspace{0.5cm}\begin{center}\begin{tabular}{|l|l|l|l|l|}
\cline{2-5}\multicolumn{1}{c|}{~}&\multicolumn{2}{c|}{{\bf$\begin{array}{c}\text{mean time }(10^{-3} \text{s) per}\\\text{helicity component}\end{array}$}}&\multicolumn{2}{c|}{{\bf total time ($10^{-3}$ s)}}\\\cline{2-5}\multicolumn{1}{c|}{{\bf~}}&\multicolumn{1}{c|}{\bf {\tt bcfw}}&\multicolumn{1}{c|}{\bf {\tt GGT/S@M}}&\multicolumn{1}{c|}{\bf {\tt bcfw}}&\multicolumn{1}{c|}{\bf {\tt GGT/S@M}}\\\hline
$\mathcal{A}_{5\phantom{0}}(gg\to ggg)$&0.11&$\phantom{\gtrsim}$0.43&2.2&$\phantom{\gtrsim}$8.7\\
$\mathcal{A}_{6\phantom{0}}(gg\to gggg)$&0.12&$\phantom{\gtrsim}$7.5&6.1&$\phantom{\gtrsim}$370\\
$\mathcal{A}_{7\phantom{0}}(gg\to ggggg)$&0.14&$\phantom{\gtrsim}$30&16&$\phantom{\gtrsim}$3,300\\
$\mathcal{A}_{8\phantom{0}}(gg\to gggggg)$&0.21&$\phantom{\gtrsim}$970&49&$\phantom{\gtrsim}$230,000\\
$\mathcal{A}_{9\phantom{0}}(gg\to ggggggg)$&0.39&$\phantom{\gtrsim}$7,300&190&$\phantom{\gtrsim}$3,600,000\\
$\mathcal{A}_{10}(gg\to gggggggg)$&1.1&$\gtrsim$1,300,000&1,100&$\gtrsim$1,300,000,000\\
$\mathcal{A}_{11}(gg\to ggggggggg)$&3.1&$\phantom{\gtrsim}$?&6,700&$\phantom{\gtrsim}$?\\\hline\end{tabular}\caption{Evaluation-times for un-polarized $n$-gluon scattering cross-sections. Where indicated, estimated times are based on extrapolation from particular helicity-amplitudes. \label{unpolarized_times}}\end{center}\vspace{-2cm}
\end{table}

\newpage

One of the functions defined by {\tt bcfw} is `{\tt Amp},' which can generate analytic formulae for any helicity-amplitude in $\mathcal{N}=4$. An example of how {\tt Amp} can be used is given in \mbox{Figure \ref{six_point_split}}.\footnote{Also used in these examples is the function `{\tt nice}' which formats formulae generated by {\tt bcfw} to be more readable---for example, by converting `{\tt ab[1,2]}'$\mapsto$`$\ab{1\,2}$'. } Using `{\tt m}' and `{\tt p}' to denote each {\tt m}inus-helicity and {\tt p}lus-helicity gluon, respectively, {\tt Amp} will generate any purely gluonic N$^k$MHV amplitude. For amplitudes involving 2 gluinos together with any number of gluons, a similar, simplified notation can be used,\footnote{An overall sign for these amplitudes has been implicitly fixed by the convention that the particle labelled `{\tt m/2}' has $SU_4$ $R$-charge $(123)$; refer to \mbox{Table \ref{R_charges_intro}}.} where `{\tt m/2}' and `{\tt p/2}' indicate the two gluinos; an example of this is given in \mbox{Figure \ref{six_point_2}}. (The reader will notice that---unless `{\tt toSpinorHelicity[n]}' is used---the only two kinematical invariants used by {\tt bcfw} are the momentum-twistor `four-bracket' \egFourBracket~and its associated `two-bracket' \egTwoBracket; these will be reviewed along with the spinor-helicity invariants in section \ref{kinematics_section}.)
\begin{figure}[t!]\begin{center}\caption{\label{six_point_split}$\mathcal{A}_{6}^{(3)}\left(-,-,-,+,+,+\right)$. The split-helicity $6$-point NMHV amplitude.\hspace{1.64cm}\vspace{-0.75cm}}\vspace{-0.25cm}\mbox{\hspace{-0.1225\textwidth}\boxed{\begin{minipage}{1.21\textwidth}\begin{tabular}{lp{0.9\textwidth}}{\color{paper_blue}{\scriptsize{\tt In[1]:=}}}&{\tt Amp\big[m,m,m,p,p,p\big]//nice}\\{\color{paper_blue}{\scriptsize{\tt In[2]:=}}}&{\tt Amp\big[m,m,m,p,p,p\big]//toSpinorHelicity[6]//nice}\\[0.05cm]{\color{paper_blue}{\scriptsize{\tt Out[1]:=}}}&{\small$\displaystyle\frac{\ab{1\,2}^3\ab{2\,3}^3\ab{3\,4\,5\,1}^3}{\ab{3\,4}\ab{4\,5}\ab{5\,6}\ab{6\,1}\ab{1\,2\,3\,4}\ab{2\,3\,4\,5}\ab{4\,5\,1\,2}\ab{5\,1\,2\,3}}+\frac{\ab{1\,2}^3\ab{2\,3}^3\ab{3\,5\,6\,1}^3}{\ab{3\,4}\ab{4\,5}\ab{5\,6}\ab{6\,1}\ab{1\,2\,3\,5}\ab{2\,3\,5\,6}\ab{5\,6\,1\,2}\ab{6\,1\,2\,3}}$}\\[0.3cm]{\color{paper_blue}{\scriptsize{\tt Out[2]:=}}}&{\small$\displaystyle-\frac{\ab{2\,3}^2\ab{3\,4}\ab{1|x_{63}x_{34}|5}^3}{\ab{4\,5}^3\ab{5\,6}\ab{6\,1}\ab{5|x_{41}x_{12}|3}s_{23}s_{34}s_{234}}-\frac{\ab{1\,2}\ab{2\,3}\ab{3|x_{25}x_{56}|1}^3}{\ab{3\,4}\ab{4\,5}\ab{6\,1}^2\ab{5|x_{41}x_{12}|3}s_{61}s_{12}s_{612}}$}\end{tabular}\end{minipage}}}\end{center}\vspace{-1cm}\end{figure}
\begin{figure}[t!]\begin{center}\caption{\label{six_point_2}$\mathcal{A}_{6}^{(3)}\left(-,-,\psi_{-1/2}^{(123)},+,+,\psi_{+1/2}^{(4)}\right)$. A $6$-point NMHV amplitude involving two gluinos and four gluons.  \vspace{-0.75cm}}
\mathematica{1}{{\small Amp\big[m,m,m/2,p,p,p/2\big]//nice}}{\mbox{\vspace{-0.15cm}$\hspace{-0.4cm}\displaystyle\displaystyle\frac{\ab{1\,2}^3\ab{2\,3}^2\ab{3\,4\,5\,1}^2}{\ab{3\,4}\ab{4\,5}\ab{5\,6}\ab{1\,2\,3\,4}\ab{5\,1\,2\,3}\ab{4\,5\,1\,2}}+\frac{\ab{1\,2}^3\ab{2\,3}^2\ab{3\,5\,6\,1}^2}{\ab{3\,4}\ab{4\,5}\ab{5\,6}\ab{1\,2\,3\,5}\ab{5\,6\,1\,2}\ab{6\,1\,2\,3}}$}}
\end{center}\vspace{-1.0cm}\end{figure}
\begin{table}[h!]\begin{center}\vspace{-0.3cm}{\small\begin{tabular}{|l|l@{\hspace{1cm}}|l|}
\hline field & $SU_4$ $R$-charge& short-notation\\\hline
$g^{}_+$&${\tt \{\}}$&$ {\tt p}$\\
$\psi_{+1/2}^{(i)}$&${\tt \{i\}}$&${\tt p/2}(\Longleftrightarrow{\tt \{4\}})$\\
$s_{0}^{(ij)}$&${\tt \{i,j\}}$&---\\
$\psi_{-1/2}^{(ijk)}$&${\tt \{i,j,k\}}$&${\tt m/2}(\Longleftrightarrow{\tt \{1,2,3\}})$\\
$g_{-}^{}$&${\tt \{1,2,3,4\}}$&${\tt m}$\\\hline
\end{tabular}}\caption{Conventions for the arguments of the functions {\tt Amp}, {\tt nAmp}, {\tt nAmpTerms}, etc.\label{R_charges_intro}}\end{center}\vspace{-0.65cm}
\end{table}

For amplitudes involving more than two gluinos (or any number of squarks), simple labels such as `{\tt m}' or `{\tt p/2}' are not sufficiently precise. This is remedied by choosing instead to label each external particle by its $SU_4$ $R$-charge, where each of the external superfields are decomposed according to\eq{\Phi^+=g^{}_+\,+\,\tilde\eta_i\,\psi_{+1/2}^{(i)}\,+\,\tilde\eta_i\tilde\eta_j\,{\phi}^{(ij)}\,+\,\tilde\eta_i\tilde\eta_j\tilde\eta_k\,\psi_{-1/2}^{(ijk)}\,+\tilde\eta_1\tilde\eta_2\tilde\eta_3\tilde\eta_4\,g_{-}^{}\;.\label{superfield}}
The syntactical rules which follow from these conventions are summarized in \mbox{Table \ref{R_charges_intro}}, but we hope they are sufficiently intuitive to be clear by example. Examples of how these more general helicity-component amplitudes can be specified are given in \mbox{Figure \ref{eight_point_example}}, which shows an 8-point N$^2$MHV helicity-amplitude involving 6 gluinos and 2 squarks, and \mbox{Figure \ref{all_fermion_10_pt}}, which shows a 10-point N$^3$MHV amplitude involving 10 gluinos. These examples also illustrate the general-purpose function `{\tt twistorSimplify},' which can often greatly simplify momentum-twistor formulae. 
\begin{figure}[t!]\begin{center}\caption{\label{eight_point_example}$\mathcal{A}_{8}^{(4)}\left(\psi_{+1/2}^{(1)},\psi_{+1/2}^{(1)},\psi_{+1/2}^{(1)},\phi_0^{(13)},\psi_{-1/2}^{(234)},\psi_{-1/2}^{(234)},\psi_{-1/2}^{(234)},\phi_0^{(24)}\right).$\hspace{2.75cm}~$~$ An example 8-point N$^2$MHV helicity-amplitude involving 6 gluinos and 2 squarks.\vspace{-0.5cm}}\vspace{0.2cm}
\mbox{\hspace{-0.5cm}\mathematica{1.05}{{\small Amp\big[\{1\},\{1\},\{1\},\{1,3\},\{2,3,4\},\{2,3,4\},\{2,3,4\},\{2\,4\}\big]; ${\%}$//twistorSimplify//nice}}{$\displaystyle\frac{\ab{5\,6}^2\ab{6\,7}^2\ab{1\,2\,3\,6}\ab{2\,3\,4\,5}}{\ab{8\,1}\ab{1\,2\,6\,7}\ab{2\,3\,5\,6}\ab{2\,3\,6\,7}\ab{3\,4\,5\,6}}$}
}\end{center}\vspace{-1.05cm}\end{figure}
\begin{figure}[t!]\begin{center}\caption{\label{all_fermion_10_pt}$\mathcal{A}_{10}^{(5)}\left(\psi_{+1/2}^{(1)},\psi_{+1/2}^{(1)},\psi_{+1/2}^{(1)},\psi_{+1/2}^{(1)},\psi_{-1/2}^{(123)},\psi_{-1/2}^{(234)},\psi_{-1/2}^{(234)},\psi_{-1/2}^{(234)},\psi_{-1/2}^{(234)},\psi_{+1/2}^{(4)}\right). $\hspace{0.5cm} $~$ An example 10-point N$^3$MHV helicity-amplitude involving only gluinos.\vspace{-0.5cm}}\vspace{0.2cm}
\mbox{\hspace{-0.5cm}\mathematica{1.05}{{\small Amp\big[\{1\},\{1\},\{1\},\{1\},\{1,2,3\},\{2,3,4\},\{2,3,4\},\{2,3,4\},\{2,3,4\},\{4\}\big]; ${\%}$//twistorSimplify//nice}}{$\displaystyle\frac{\ab{5\,6}\ab{6\,7}^2\ab{7\,8}^2\ab{8\,9}^2\ab{1\,2\,3\,9}\ab{2\,3\,4\,8}\ab{3\,4\,5\,7}}{\ab{10\,\,1}\ab{1\,2\,8\,9}\ab{2\,3\,7\,8}\ab{2\,3\,8\,9}\ab{3\,4\,6\,7}\ab{3\,4\,7\,8}\ab{4\,5\,6\,7}}$}
}\end{center}\vspace{-1.05cm}\end{figure}

This paper is outlined as follows. In the next section we will review the kinematics of momentum-twistors and their connection to ordinary four-momenta and spinor-helicity variables. In \mbox{section \ref{grassmannian_tree_contours}}, we review the tree-level BCFW recursion-relations as a statement about contour integrals in the momentum-twistor Grassmannian, \cite{ArkaniHamed:2009vw,Mason:2009qx}, and describe a three-parameter family of recursive `schemes' in which the BCFW recursion relations can be implemented. In section \ref{bcfw_package_section} we describe the basic use of the {\tt bcfw} package along with its principle  functions. (A more thorough walk-through, containing numerous example computations, can be found in the {\sc Mathematica} notebook \mbox{{\tt bcfw-v0-walk-through.nb}} distributed alongside the {\tt bcfw} package---attached to the submission file for this note to the {\tt arXiv}.) In section \ref{comparison_section} we briefly discuss {\tt bcfw} in the context of other existing computational tools including the {\sc Mathematica} package {\tt Gluon-Gluino-Trees} ({\tt GGT}), \cite{Dixon:2010ik}. In \mbox{appendix \ref{glossary_of_functions}} we include an index of the key functions which are made available by the package {\tt bcfw}.

\vspace{-0.5cm}\section{Kinematics: Momenta to Momentum-Twistors (and Back)}\label{kinematics_section}
By default, all tree-amplitudes generated by the {\tt bcfw} package are handled internally as purely-holomorphic functions of the momentum-twistor variables $\{Z_a\}$ introduced by Andrew Hodges in \cite{Hodges:2009hk}, together with an overall MHV-amplitude pre-factor which also depends on what is known as the `infinity (bi-)twistor,' $I_{\infty},$ which associates with each momentum-twistor $Z_a$ a Lorentz spinor $\lambda_a^{\underline{\alpha}=1,2}$ in the fundamental representation of $SL_2(\mathbb{C})$. In addition to the many theoretical advantages of working with momentum-twistors, there are many indications that tree amplitudes are most compactly-written and most efficiently-evaluated in terms of momentum-twistors. But before we review this relatively novel formalism, we should reiterate that {\tt bcfw} is fully-equipped to work with kinematics specified in terms of four-momenta or spinor-helicity variables (or momentum-twistors, of course), and can convert momentum-twistor formulae into those involving spinor-helicity variables and dual coordinates (but at a substantial cost in efficiency). Because of this, {\tt bcfw} should be relatively easy to incorporate into other computational frameworks. 

The connection between ordinary four-momenta $p^{\mu}$ and momentum-twistors starts with the association of a (Hermitian) matrix $p^{\underline{\alpha}\,\dot{\underline{\alpha}}}$ with each (real) four-momentum \nolinebreak $p^\mu$,
\eq{p^\mu\mapsto p^{\alpha\dot{\alpha}}\equiv p^\mu\sigma_{\mu}^{\alpha\dot{\alpha}}=\left(\begin{array}{cc}p_0+p_3&p_1-ip_2\\p_1+ip_2&p_0-p_3\end{array}\right).}
Noticing that $p^{\mu}p_{\mu}=\det(p^{\underline{\alpha}\,\dot{\underline{\alpha}}})$, it follows that light-like momenta are represented by matrices with vanishing determinant. Any such matrix can be written as an outer-product, \eq{\det(p^{\underline{\alpha}\,\dot{\underline{\alpha}}})=0\quad\Longleftrightarrow\quad p^{\underline{\alpha}\,\dot{\underline{\alpha}}}\equiv\lambda^{\underline{\alpha}}\tilde{\lambda}^{\dot{\underline{\alpha}}},} where $\lambda$ and $\tilde\lambda$ are the famous {\it spinor-helicity} variables. For real momenta, it is easy to see that $\tilde\lambda^{\dot{\underline\alpha}}=\pm\left(\lambda^{\underline\alpha}\right)^{*}$, where the sign is determined by whether $p^{\mu}$ has positive or negative energy, respectively. Of course, this identification is only defined up-to an arbitrary phase: $\lambda\mapsto e^{i\theta}\lambda, \tilde\lambda\mapsto e^{-i\theta}\tilde\lambda$. Such re-phasing is induced by the action of little-group for massless particles in four-dimensions. 

One of the principle advantages to working with spinor-helicity variables is that any function built out of the $SL_2(\mathbb{C})$-invariants \eqs{\hspace{-4cm}&\ab{\lambda_a\,\lambda_b}\equiv\ab{ab}\equiv\det(\lambda_a\lambda_b)=\left|\begin{array}{cc}\lambda_a^{\underline{1}}&\lambda_b^{\underline{1}}\\\lambda_a^{\underline{2}}&\lambda_b^{\underline{2}}\end{array}\right|,\\\hspace{-2cm}\mathrm{and}\qquad&\,[\tilde\lambda_a\,\tilde\lambda_b]\,\equiv\left[ab\right]\equiv\det(\tilde\lambda_a\,\tilde\lambda_b)=\left|\begin{array}{cc}\tilde{\lambda}_a^{\dot{\underline{1}}}&\tilde{\lambda}_b^{\dot{\underline{1}}}\\\tilde{\lambda}_a^{\dot{\underline{2}}}&\tilde{\lambda}_b^{\dot{\underline{2}}}\end{array}\right|,} will automatically be Lorentz-invariant up to little-group re-phasing. Amplitudes involving massless particles, therefore, when written in terms of spinor-helicity variables, will be functions with uniform weight under $\lambda_a\mapsto u\lambda_a$ (with weight equal to minus twice the helicity of particle $a$).

The next step along the road from momenta to momentum-twistors are {\it dual coordinates} $x_a^{\underline{\alpha}\,\dot{\underline{\alpha}}}$ (also known as {\it region momenta}) defined (implicitly) through the identification \eq{p_{a}\equiv x_{a}-x_{a-1}.}
(Whenever it is necessary to fix a convention, we will choose $x_1$ to be the origin of dual coordinate space.) One of the most important recent discoveries regarding scattering amplitudes in $\mathcal{N}=4$ SYM is that, after diving by the $n$-point MHV tree-amplitude, scattering amplitudes in $\mathcal{N}=4$ are not just superconformally-invariant in ordinary spacetime, but are also superconformally-invariant with respect to these dual-coordinates, \cite{Berkovits:2008ic,Drummond:2008vq}, and this is made manifest term-by-term in BCFW, \cite{Drummond:2008cr}. The existence of a conformal symmetry on this dual space led Andrew Hodges to propose in \cite{Hodges:2009hk} that amplitudes be described in the twistor-space associated with these dual coordinates; the twistor space of dual-coordinates is known as {\it momentum twistor} space. 
\begin{figure}[h]\vspace{-0.3cm}\begin{center}\caption{The map connecting momentum-twistor variables and dual-coordinates.\label{momentum_twistor_geometry}}\includegraphics[scale=0.8]{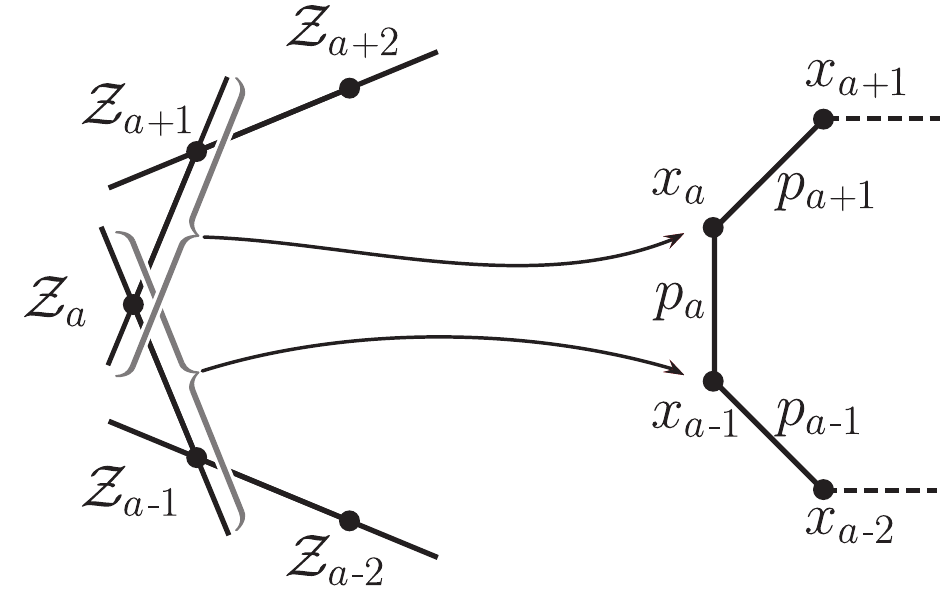}\vspace{0cm}\end{center}\vspace{-0.8cm}\end{figure}

Because each pair of {\it consecutive} dual coordinates are null-separated (the momenta being on-shell), the null-line joining them corresponds to a single momentum-twistor. And because the pair of dual coordinates $(x_a,x_{a-1})$ encode the null-momentum $p_a$, it is natural to call the momentum-twistor associated with this null-line `$Z_a$'. Making this identification will associate the line $(Z_a,Z_{a-1})$ in momentum-twistor space with the point $x_{a-1}$, and the line $(Z_{a+1},Z_a)$ with the point $x_a$; that these two lines intersect at the twistor $Z_a$ reflects the fact that the points $x_a$ and $x_{a-1}$ are null-separated. 

Using the conventions just established, we canonically associate a momentum-twistor $Z_a$ to each momentum $p_a$ according to the rule,
\eq{p_a=\lambda_a\tilde\lambda_a=x_{a}-x_{a-1}\qquad\Longleftrightarrow\qquad Z_a\equiv\left(\begin{array}{c}\lambda_a^{\underline{\alpha}}\\x^{\phantom{\alpha}\phantom{a}\dot{\underline{\alpha}}}_{a\,\underline{\alpha}}\lambda_a^{\underline{\alpha}}\end{array}\right)\equiv\left(\begin{array}{c}\lambda_a^{\underline{\alpha}}\\\mu_a^{\underline{\dot{\alpha}}}\end{array}\right)\label{momentum_twistor_x_def}.} 
Notice that our convention of choosing $x_1$ as the origin of dual-coordinate space trivially fixes $\mu_1$={\scriptsize$\left(\begin{array}{c}0\\0\end{array}\right)$}. Moreover, because this implies that \mbox{$p_2=\lambda_2\tilde\lambda_2=x_2-x_1=x_2$}, we see that \mbox{$\mu_2=x^{\phantom{\alpha}\phantom{2}\dot{\underline{\alpha}}}_{2\,\underline{\alpha}}\lambda_2^{\underline{\alpha}}(\propto\ab{\lambda_2\,\lambda_2})=${\scriptsize$\left(\begin{array}{c}0\\0\end{array}\right)$}} as well. Working out the rest of this map explicitly---as was described in \cite{ArkaniHamed:2009sx}---we find that we may write
{\small \eq{\mu_a=\qabinverse_{ab}\tilde{\lambda}_b,\qquad\mathrm{where}\qquad\qabinverse_{ab}=\left(\begin{array}{ccccccccc}
0&0&0&\cdots&\cdots&\cdots&0\\
0&0&0&0&\cdots&\cdots&0\\
0&\ab{2\,3}&0&0&\ddots&\ddots&0\\
0&\ab{2\,4}&\ab{3\,4}&0&0&\ddots&0\\
0&\ab{2\,5}&\ab{3\,5}&\ab{4\,5}&0&\ddots&0\\
\vdots&\vdots&\vdots&\vdots&\ddots&\ddots&0\\
0&\ab{2\,n}&\ab{3\,n}&\ab{4\,n}&\cdots&\ab{n-1\,n}&0\end{array}\right).\label{qInverse_definition}}}
$\!\!\qabinverse_{ab}$ is so-named because it is a `Formal-inverse' of the (singular) map $\qab_{ab}$ which relates the $\mu$'s to the $\tilde\lambda$'s according to $\tilde\lambda_a=\qab_{ab}\mu_b$ where 
{\small\eq{\qab_{ab}=\left(\begin{array}{ccccccccc}
\frac{\ab{2\,n}}{\ab{n\,1}\ab{1\,2}}&\frac{1}{\ab{1\,2}}&0&\cdots&\cdots&\cdots&\frac{1}{\ab{n\,1}}\\
\frac{1}{\ab{1\,2}}&\frac{\ab{3\,1}}{\ab{1\,2}\ab{2\,3}}&\frac{1}{\ab{2\,3}}&0&\cdots&\cdots&0\\
0&\frac{1}{\ab{2\,3}}&\frac{\ab{4\,2}}{\ab{2\,3}\ab{3\,4}}&\frac{1}{\ab{3\,4}}&0&\ddots&\vdots\\
\vdots&0&\frac{1}{\ab{3\,4}}&\frac{\ab{5\,3}}{\ab{3\,4}\ab{4\,5}}&\frac{1}{\ab{4\,5}}&\ddots&0\\
0&\vdots&\ddots&\ddots&\ddots&\ddots&\frac{1}{\ab{n-1\,n}}\\
\frac{1}{\ab{n\,1}}&0&\cdots&\cdots&0&\frac{1}{\ab{n-1\,n}}&\frac{\ab{1\,n-1}}{\ab{n-1\,n}\ab{n\,1}}\end{array}\right).\label{qab_definition}}}
\noindent It is worth emphasizing that although $\qab_{ab}$ is singular, our conventions ensure that $\mu_a=\qabinverse_{ab}\qab_{bc}\mu_c,$ and $\tilde\lambda_a=\qab_{ab}\qabinverse_{bc}\tilde\lambda_{c},$ which justifies calling $\qabinverse_{ab}$ the `inverse' of $\qab_{ab}$. 

What we have described so far have been ordinary (Bosonic) momentum twistors; these have a natural extension to momentum-{\it super}twistors defined by 
\eq{\mathcal{Z}_a\equiv\left(\begin{array}{c}Z_a\\\eta_a\end{array}\right)\equiv\left(\begin{array}{c}\lambda_a\\\mu_a\\\eta_a\end{array}\right),}
where the Fermionic $\eta$-components of the supertwistors are related to the ordinary Fermionic parameters $\tilde\eta$ which define each superfield (\ref{superfield}) in precisely the same way that the $\mu$ variables are related to the $\tilde\lambda$ variables. To summarize, the components of the momentum-supertwistors are related to the ordinary spinor-helicity variables via \eq{\!\!\!\!\!\lambda_a^{\underline{\alpha}=1,2}=Z_a^{1,2},\!\quad\mathrm{and}\qquad\!\mu_a^{\underline{\dot{\alpha}}=1,2}=Z_a^{3,4},}\eq{\tilde{\lambda}_a= \qab_{ab}\mu_b,\qquad\mathrm{and}\qquad\mu_a=\qabinverse_{ab}\tilde{\lambda}_b,}
\eq{\tilde{\eta}_a= \qab_{ab}\eta_b,\qquad\mathrm{and}\qquad\eta_a=\qabinverse_{ab}\tilde{\eta}_b.}

Just as spinor-helicity variables went a long way toward trivializing Lorentz-invariance, momentum-twistors essentially trivialize momentum conservation and dual conformal invariance. Momentum conservation is trivial because {\it any} set of $n$ (ordered) momentum twistors will define $n$ null-separated region momenta through the maps given above. Furthermore, up to little-group rescaling, dual-conformal transformations act on momentum-twistors as $SL_4(\mathbb{C})$ transformations, meaning that any function of the (only) natural $SL_4(\mathbb{C})$-invariant product---namely, `$\det$'---will automatically be dual-conformally invariant if it has appropriate little-group weights. This suggests the natural generalization of the `angle-bracket' $\ab{a\,b}$ defined for 2-spinors above would be the momentum-twistor four-bracket \egFourBracket~defined according to 
\eq{\hspace{0.5cm}{\tt ab[a,b,c,d]}\Longleftrightarrow \ab{a\,b\,c\,d}\equiv\left|\begin{array}{cccc}Z_a^1&Z_b^1&Z_c^1&Z_d^1\\Z_a^2&Z_b^2&Z_c^2&Z_d^2\\Z_a^3&Z_b^3&Z_c^3&Z_d^3\\Z_a^4&Z_b^4&Z_c^4&Z_d^4\end{array}\right|\Longleftrightarrow{\tt Det[Zs[[\{a,b,c,d\}]]]};}

So it would appear that, including also the MHV-amplitude pre-factor, all amplitudes can be written in terms of four-brackets \egFourBracket ~and two-brackets \egTwoBracket; but it is easy to see that the latter is just a special-case of the former. Notice that the map connecting a momentum-twistor $Z_a$ and ordinary spinor-helicity variables, \mbox{equation (\ref{momentum_twistor_x_def})}, is a {\it component-wise} definition. Because any such definition is manifestly {\it not} $SL_4(\mathbb{C})$-invariant, this map breaks dual-conformal invariance. We can make this clear by choosing to write $I_{\infty}$ explicitly, defining two-brackets via,\eq{{\tt ab[a,b]}\Longleftrightarrow\ab{a\,b}\equiv\ab{a\,b\,I_{\infty}}\equiv\left|\begin{array}{cccc}Z_a^1&Z_b^1&0&0\\Z_a^2&Z_b^2&0&0\\Z_a^3&Z_b^3&1&0\\Z_a^4&Z_b^4&0&1\end{array}\right|\Longleftrightarrow{\tt Det[Zs[[\{a,b\},1;;2]]]}.}

Because momentum twistors are still somewhat unfamiliar to many researchers, we should mention that there is a completely canonical map between four-brackets and ordinary spinor-helicity variables which follows directly from definition (\ref{momentum_twistor_x_def}). Rather than giving this map for a completely general four-bracket, we will see in the next section that tree-level BCFW only generates formulae involving four-brackets which involve at least one pair of adjacent momentum-twistors---that is, tree amplitudes involve only four-brackets of the form $\ab{a\,\,j\,\,j\smallplus1\,\,b}$. Using (\ref{momentum_twistor_x_def}), it is easy to see that \eq{\ab{a\,\,j\,\,j\smallplus1\,\,b}=\ab{j\smallplus1\,\,j}\ab{a|x_{a\,j}x_{j\,b}|b},}
where we have used the notation $x_{a\,\,b}\equiv x_{b}-x_a$.\footnote{This notation (and sign-convention) becomes clearer if $x_{a\,\,b}$ is viewed as the vector from $x_{a}$\nolinebreak to \nolinebreak $x_{b}$.} This further simplifies in the special case of a four-bracket involving two pairs of adjacent momentum-twistors, \eqs{\ab{a\smallminus1\,\,a\,\,b\,\,b\smallplus1}&=\ab{a\smallminus1\,\,a}\ab{b\,\,b\smallplus1}(p_a+p_{a+1}+\ldots+p_{b-1}+p_b)^2\\&\equiv\ab{a\smallminus1\,\,a}\ab{b\,\,b\smallplus1}s_{a\cdots b}\equiv\ab{a\smallminus1\,\,a}\ab{b\,\,b\smallplus1}x_{a-1\,\,b}^2.}

It is worth mentioning that the fact that tree-level BCFW involves only four-brackets of the form $\ab{a\,\,j\,\,j\smallplus1\,\,b}$ means that in general, every superamplitude in \mbox{$\mathcal{N}=4$} involves strictly fewer than {\scriptsize$\left(\begin{array}{c}n\\4\end{array}\right)$} kinematical invariants.

\newpage
\section{\mbox{Tree-Amplitudes as Contour Integrals in the Grassmannian}}
\begin{figure}[h!]\begin{center}\vspace{-0.5cm}\includegraphics[scale=0.7]{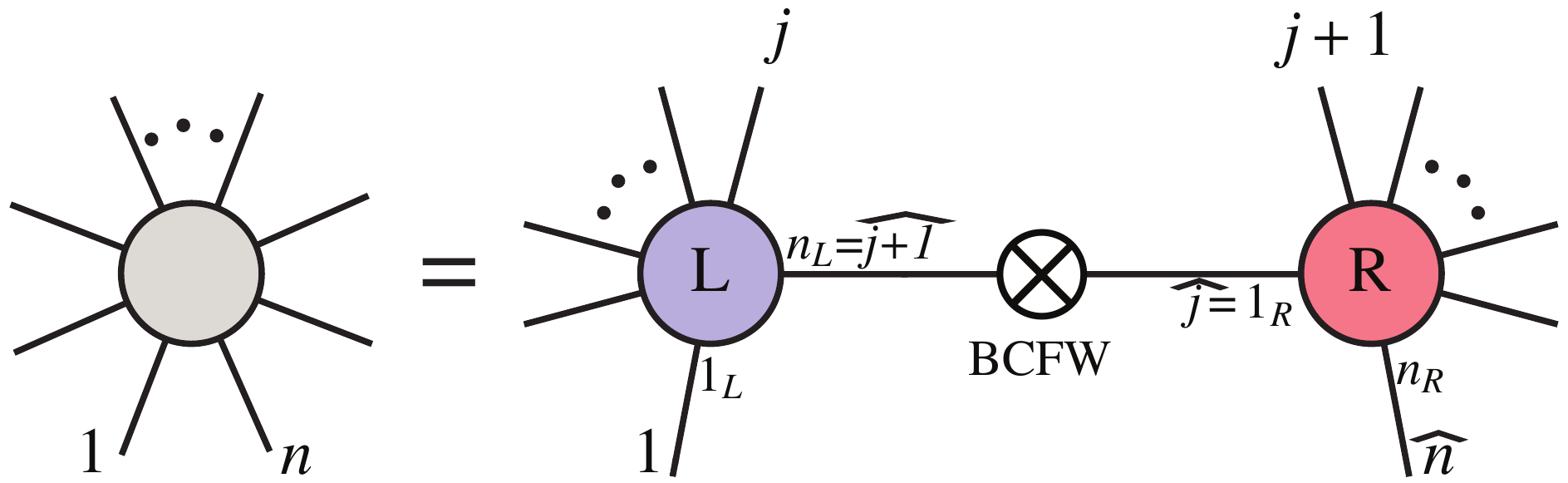}\caption{The momentum-twistor BCFW-bridge (without any rotations).\label{classic_bridge}}\end{center}\vspace{-1.05cm}\end{figure}

\vspace{-0.4cm}\label{grassmannian_tree_contours}
The {\tt bcfw} package describes each $n$-point N$^k$MHV tree-amplitude as a contour integral in the Grassmannian $G(k,n)$ of $k$-planes in $n$-dimensions (see \cite{ArkaniHamed:2010kv,ArkaniHamed:2009dg,ArkaniHamed:2009dn,ArkaniHamed:2009vw}),\vspace{-0.025cm}
\eqs{\label{grassmannian_contour}\mathscr{A}_{n}^{(m=k+2)}&=\frac{1}{\mathrm{vol}(GL_k)}\!\oint\limits_{\Gamma_{n,m}}\!\!\frac{d^{n\times k} D_{\alpha\,a}\,\prod_{\alpha=1}^k\delta^{4|4}\left(D_{\alpha\,a}\mathcal{Z}_a\right)}{(1\cdots k)(2\cdots k\smallplus1)\cdots(n\cdots k\smallminus1)},\\&=\sum_{\gamma\in\Gamma_{n,m}}\left\{\frac{1}{\mathrm{vol}(GL_k)}\!\!\!\!\!\!\!\!\!\!\!\!\oint\limits_{|D_{\alpha\,a}-({\tt dMatrix}_\gamma)|=\epsilon}\!\!\!\!\!\!\!\!\!\!\!\!\frac{d^{n\times k} D_{\alpha\,a}\,\prod_{\alpha=1}^k\delta^{4|4}\left(D_{\alpha\,a}\mathcal{Z}_a\right)}{(1\cdots k)(2\cdots k\smallplus1)\cdots(n\cdots k\smallminus1)}\right\},\\&=\sum_{\gamma\in\Gamma_{n,m}}\left\{({\tt residue}_{\gamma})\prod_{\alpha=1}^{k}\delta^{0|4}\Big(({\tt dMatrix}_\gamma)_{\alpha\,a}\eta_a\Big)\right\},\vspace{-0.05cm}} where we have used the scripted `$\mathscr{A}_n^{(m)}$' to indicate that this is the tree-amplitude {\it divided by the (supersymmetric)} $n$-point MHV-amplitude\footnote{Here, we are not including the ordinary momentum-conserving $\delta$-function, $\delta^4(\lambda_a\tilde\lambda_a)$, because all momentum-twistor amplitudes are automatically on its support.}, \eq{\mathcal{A}_n^{(2)}=\frac{\prod_{\underline{\alpha}=1}^2\delta^{0|4}\big(\lambda_{a}^{\underline{\alpha}}\tilde\eta_a\big)}{\ab{1\,2}\ab{2\,3}\cdots\ab{n\smallminus1\,n}\ab{n\,1}}.}
As all the terms generated by the BCFW recursion relations are Yangian-invariant \cite{Drummond:2009fd}, they are each residues of the integral (\ref{grassmannian_contour}), \cite{Drummond:2010qh,Drummond:2010uq}---computed for contours which `encircle' isolated poles in the Grassmannian . Therefore, each term can be described as a part of the complete `tree-contour' $\Gamma_{n,m}$. This helps to explain the nomenclature of {\tt bcfw}, where each superamplitude stored as a function called `{\tt treeContour}.' Notice that the coefficients appearing in the Fermionic $\delta$-functions of (\ref{grassmannian_contour}), ${\tt dMatrix}_{\gamma},$ directly represent the isolated {\it points} in $G(k,n)$ where the integral (\ref{grassmannian_contour}) develops a pole (of the appropriate co-dimension) which is to be `encircled' by the contour $\Gamma_{n,m}$, each giving rise to a particular residue of the integral. Of course, knowing the poles---that is, knowing {\it just} the list of points in $G(k,n)$ (and the orientation of the contour about each)---is sufficient to calculate each residue using the contour integral (\ref{grassmannian_contour}); but it turns out that this is in fact unnecessary for our purposes: the BCFW recursion relations directly calculate the {\it residues} themselves in a canonical way. 
\newpage
As described in \cite{ArkaniHamed:2010kv}, when expressed in terms of momentum-twistor variables, the tree-level BCFW recursion relations become the following. 
\vspace{-0.15cm}\begin{align}\label{bcfw_in_momentum_twistors}\vspace{-1cm}\hspace{-1cm}\mathscr{A}_n^{(m=k+2)}=&\phantom{\,+\,} \mathscr{A}_{n-1}^{(m)}\\\hspace{-1cm}&+\sum_{\substack{n_L,m_L\\n_R,m_R}} \mathscr{A}_{n_L}^{(m_L)}(1,\ldots,j,\hat{j\smallplus1})R[n\smallminus1\,n\,1\,j\,j\smallplus1] \mathscr{A}_{n_R}^{(m_R)}(\hat{j},j\smallplus1,\ldots,n\smallminus1,\hat{n}),\nonumber\end{align}\vspace{-0.0cm}
where,\footnote{It is worth noting that $\widehat{\mathcal{Z}_{j+1}}$ and $\widehat{\mathcal{Z}_{j}}$ are projectively equivalent; the reason for distinguishing them as in (\ref{zHats}) is to preserve canonical little-group assignments.} \vspace{-0.0cm}
{\normalsize\eqs{\hat{\mathcal{Z}_{j+1}}&=(j\smallplus1\,j)\cap(n\smallminus1\,n\,1)\equiv\mathcal{Z}_{j+1}+\mathcal{Z}_{j}\frac{\ab{j\smallplus1\,n\smallminus1\,n\,1}}{\ab{n\smallminus1\,n\,1\,j}},\\\hat{\mathcal{Z}_{j}}&=(j\,j\smallplus1)\cap(n\smallminus1\,n\,1)\equiv\mathcal{Z}_j+\mathcal{Z}_{j+1}\frac{\ab{j\,n\smallminus1\,n\,1}}{\ab{n\smallminus1\,n\,1\,j\smallplus1}},\\\widehat{\mathcal{Z}_n}&=(n\,n\smallminus1)\cap(1\,j\,j\smallplus1)\equiv\mathcal{Z}_n+\mathcal{Z}_{n-1}\frac{\ab{n\,1\,j\,j\smallplus1}}{\ab{1\,j\,j\smallplus1\,n\smallminus1}},\label{zHats}}}\vspace{-0.25cm}
and
\vspace{-0.0cm}\eq{\hspace{-0.35cm}R[a\,b\,c\,d\,e]\equiv\frac{\delta^{0|4}\Big(\eta_a\ab{b\,c\,d\,e}+\eta_b\ab{c\,d\,e\,a}+\eta_c\ab{d\,e\,a\,b}+\eta_d\ab{e\,a\,b\,c}+\eta_e\ab{a\,b\,c\,d}\Big)}{\ab{a\,b\,c\,d}\ab{b\,c\,d\,e}\ab{c\,d\,e\,a}\ab{d\,e\,a\,b}\ab{e\,a\,b\,c}}.\vspace{-0.2cm}}
This tree-level BCFW-bridge is illustrated in \mbox{Figure \ref{classic_bridge}}.

The shifted momentum-twistors in (\ref{zHats}) should be understood supersymmetrically, and the shifted Fermionic $\eta$-variables result in a shifted matrix of coefficients. Specifically, for terms bridged in the recursion, the residues (evaluated with shifted arguments) are simply multiplied, and the supersymmetric $\delta^{0|4}$'s combine according to:
\eq{\underbrace{{\color{paper_blue}\left(\begin{array}{cccccccc}d^L_{1,1}&\cdots&\cdots&\cdots&d^L_{1,n_L}\\\vdots&\vdots&\text{\LARGE $L$}&\vdots&\vdots\\d^{L}_{k_L,1}&\cdots&\cdots&\cdots&d^L_{k_L,n_L}\end{array}\right)}\underset{\mathrm{BCFW}}{\bigotimes}{\color{paper_red}\left(\begin{array}{cccccccc}d^R_{1,1}&\cdots&\cdots&\cdots&d^R_{1,n_R}\\\vdots&\vdots&\text{\LARGE $R$}&\vdots&\vdots\\d^{R}_{k_R,1}&\cdots&\cdots&\cdots&d^R_{k_R,n_R}\end{array}\right)}}_\Big\Downarrow\nonumber}
{\footnotesize\eq{\nonumber\hspace{-2.05cm}\left(\!\!\!\!\begin{array}{ccccccccccc}
{\color{paper_blue}\raisebox{-0.65cm}[0cm][-1cm]{$\!\!\!\!\!\left(\raisebox{0.95cm}{$$}\right.$}d^L_{1,1}}&{\color{paper_blue}d^L_{1,2}}&{\color{paper_blue}\cdots}&{\color{paper_blue}d^L_{1,j\!-\!1}}&{\color{paper_blue}\left(\!d^L_{1,j}\!+\!\zeta^{L}_{j\!+\!1}d^L_{1,j\!+\!1}\!\right)}&{\color{paper_blue}d^L_{1,j+1}\raisebox{-0.65cm}[0cm][-1cm]{$\,\left.\raisebox{0.95cm}{$$}\right)\!\!\!\!\!\!\!$}}&{\color{paper_grey}0}&{\color{paper_grey}\cdots}&{\color{paper_grey}0}&{\color{paper_grey}0}&{\color{paper_grey}0}\\
{\color{paper_blue}\vdots}&{\color{paper_blue}\vdots}&{\color{paper_blue}\text{\LARGE $L$}}&{\color{paper_blue}\vdots}&{\color{paper_blue}\vdots}&{\color{paper_blue}\vdots}&{\color{paper_grey}\vdots}&{\color{paper_grey}\ddots}&{\color{paper_grey}\vdots}&{\color{paper_grey}\vdots}&{\color{paper_grey}\vdots}\\
{\color{paper_blue}d^L_{k_L,1}}&{\color{paper_blue}d^L_{k_L,2}}&{\color{paper_blue}\cdots}&{\color{paper_blue}d^L_{k_L,j\!-\!1}}&{\color{paper_blue}\left(\!d^L_{k_L,j}\!\!+\!\zeta^{L}_{j\!+\!1}d^L_{k_L,j\!+\!1}\!\right)}&{\color{paper_blue}d^L_{k_L,j\!+\!1}}&{\color{paper_grey}0}&{\color{paper_grey}\cdots}&{\color{paper_grey}0}&{\color{paper_grey}0}&{\color{paper_grey}0}\\
\ab{j\,j\smallplus1\,n\smallminus1\,n}&0&\cdots&0&\ab{j\smallplus1\,n\smallminus1\,n\,1}&\ab{n\smallminus1\,n\,1\,j}&0&\cdots&0&\ab{n\,1\,j\,j\smallplus1}&\ab{1\,j\,j\smallplus1\,n\smallminus1}\\
{\color{paper_grey}0}&{\color{paper_grey}0}&{\color{paper_grey}\cdots}&{\color{paper_grey}0}&{\color{paper_red}\raisebox{-0.65cm}[0cm][-1cm]{$\!\!\!\!\!\left(\raisebox{0.95cm}{$$}\right.$}d^R_{1,j}}&{\color{paper_red}\left(\!d^R_{1,j\!+\!1}\!+\!\zeta^{R}_jd^R_{1,j}\!\right)}&{\color{paper_red}d^R_{1,j\!+\!2}}&{\color{paper_red}\cdots}&{\color{paper_red}d^R_{1,n\!-\!2}}&{\color{paper_red}\left(\!d^R_{1,n\!-\!1}+\zeta_{n}^Rd^R_{1,n}\right)}&{\color{paper_red}d^R_{1,n}\raisebox{-0.65cm}[0cm][-1cm]{$\,\left.\raisebox{0.95cm}{$$}\right)\!\!\!\!\!\!$}}\\
{\color{paper_grey}\vdots}&{\color{paper_grey}\vdots}&{\color{paper_grey}\ddots}&{\color{paper_grey}\vdots}&{\color{paper_red}\vdots}&{\color{paper_red}\vdots}&{\color{paper_red}\vdots}&{\color{paper_red}\text{\LARGE $R$}}&{\color{paper_red}\vdots}&{\color{paper_red}\vdots}&{\color{paper_red}\vdots}\\
{\color{paper_grey}0}&{\color{paper_grey}0}&{\color{paper_grey}\cdots}&{\color{paper_grey}0}&{\color{paper_red}d^R_{k_R,j}}&{\color{paper_red}\left(\!d^R_{k_R,j\!+\!1}\!\!+\!\zeta^{R}_jd^R_{k_R,j}\!\right)}&{\color{paper_red}d^R_{k_R,j\!+\!2}}&{\color{paper_red}\cdots}&{\color{paper_red}d^R_{k_R,n\!-\!2}}&{\color{paper_red}\left(\!d^R_{k_R,n\!-\!1}\!\!+\!\zeta_{n}^Rd^R_{k_R,n}\!\right)}&{\color{paper_red}d^R_{k_R,n}}
\end{array}\!\!\!\!\right)}}
with
\eq{{\color{paper_blue}\zeta^L_{j+1}}\equiv\frac{\ab{j\smallplus1\,n\smallminus1\,n\,1}}{\ab{n\smallminus1\,n\,1\,j}},\qquad{\color{paper_red}\zeta^R_j}\equiv
\frac{\ab{j\,n\smallminus1\,n\,1}}{\ab{n\smallminus1\,n\,1\,j\smallplus1}},\quad\mathrm{and}\quad{\color{paper_red}\zeta^R_n}\equiv\frac{\ab{n\,1\,j\,j\smallplus1}}{\ab{1\,j\,j\smallplus1\,n\smallminus1}}.}
Thus, the tree-level BCFW recursion relations amount to little more than cutting-and-pasting (and re-labeling) matrices, allowing most amplitudes of interest to be recursed in essentially real-time. 

\subsection{Generalized BCFW Recursion Schemes}\label{recursion_scheme_subsection}
Although the recursive BCFW formula (\ref{bcfw_in_momentum_twistors}) fixes $\mathscr{A}_n^{(m)}$ given all amplitudes with strictly fewer particles, (\ref{bcfw_in_momentum_twistors}) by itself does not uniquely identify any {\it particular} sum of residues. The reason for this is simple (and completely trivial): the lower-point amplitudes appearing in the recursion (\ref{bcfw_in_momentum_twistors}) can be written in any way whatsoever---with many choices corresponding to all the representatives $\Gamma_{n,m}$ of each tree-contours' homology-class. Said another way, in order to use (\ref{bcfw_in_momentum_twistors}) to obtain {\it a particular} contour for the $n$-point amplitude, it is necessary to know the {\it particular, representative} contours for all lower-point amplitudes; but these lower-point contours need-not have been recursed in any particular way. In order to obtain {\it an explicit, representative contour} through the use of the BCFW recursion relations---i.e.\ using (\ref{bcfw_in_momentum_twistors})---it is necessary to give a prescription for how {\it all} lower-point amplitudes are also to be recursed.

One especially natural prescription would be to recurse {\it all} lower-point amplitudes {\it exactly} according to equation (\ref{bcfw_in_momentum_twistors})---with each $n$-point amplitude having ordered-arguments \mbox{$(1,\ldots,n)$}. This is the default recursive scheme used by {\tt bcfw} and is obtained with the function {\tt treeContour[n,m]=generalTreeContour[0,0,0][n,m]}$\!$. This scheme follows from \mbox{Figure \ref{classic_bridge}} where each lower-point amplitude is recursed precisely according to \mbox{Figure \ref{classic_bridge}}.  

\begin{figure}[h!]\begin{center}\vspace{-0.8cm}$\begin{array}{c}\includegraphics[scale=0.5]{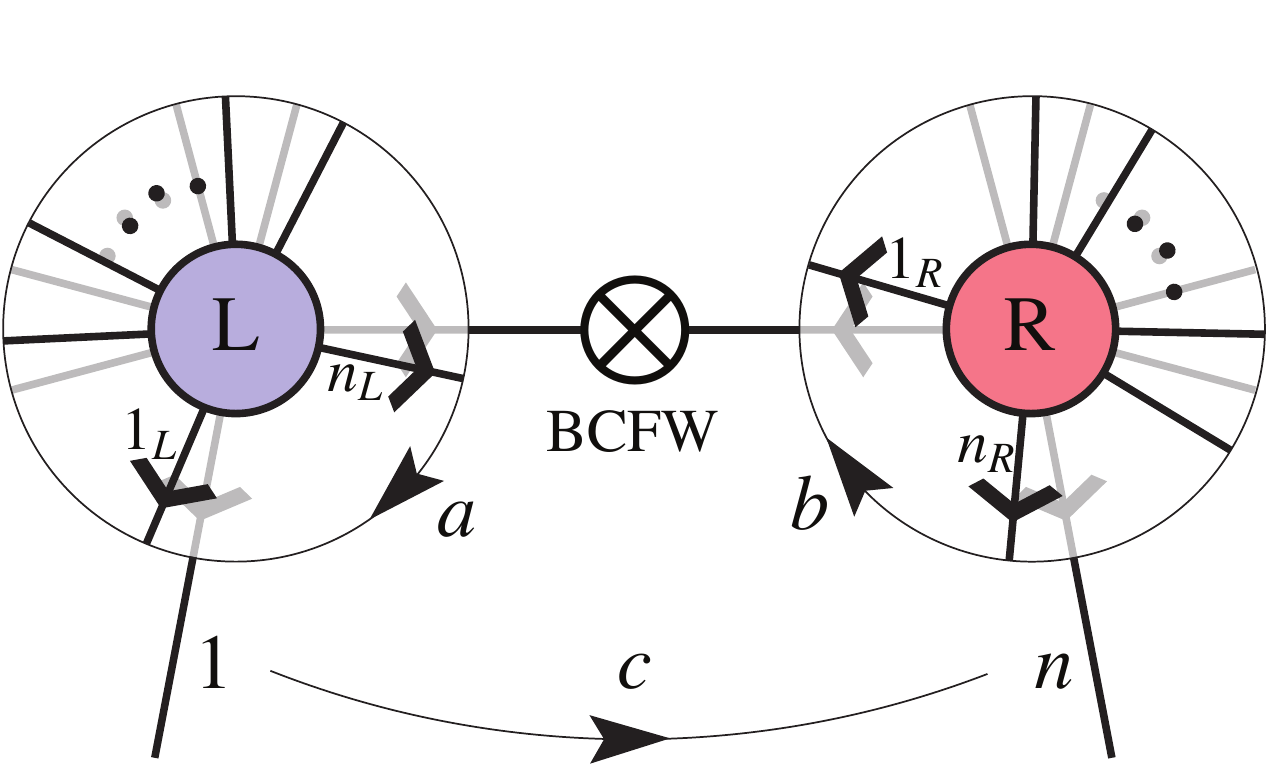}\\{\tt generalTreeContour[a,b,c]}\end{array}$\caption{An illustration of the generalized BCFW recursion-schemes used by {\tt bcfw}'s function {\tt generalTreeContour[a,b,c]}. Here, the legs being deformed in the left-hand amplitude, for example, should be thought-of as being actively `rotated' clockwise by an amount `{\tt a}' relative to the default recursive scheme.\label{general_contour_figure}}\vspace{-0.85cm}\end{center}\end{figure}

Among the many recursive prescriptions one could imagine, a remarkable degree of complexity results from simply allowing for arbitrary (and separate) `rotations' of the amplitudes appearing on the left- and right-hand sides of the BCFW bridge,\footnote{When making these rotations, the homogeneous term in the recursion, $\mathscr{A}_{n-1}^{(m)}$, must be considered an amplitude occurring on the {\it left}.} and also allowing for an over-all rotation of the the $n$-point amplitude being recursed---or equivalently, which legs are deformed in the recursion. Specifically, letting $g$ denote a cyclic-rotation of (an explicit formula) of an amplitude \mbox{$g:\mathscr{A}_n(1,\ldots,n)\mapsto \mathscr{A}_n(2,\ldots,n,1)$}; then the class of generalized BCFW recursion schemes implemented in {\tt bcfw} is given by, 
\eqs{\nonumber\vspace{-1cm}\hspace{-1.0cm}&{\tt generalTreeContour[a,b,c][n,m]:}g^{-{\tt c}}\big[\mathscr{A}_n^{(m)}\big]\underset{{\tt\{a,b,c\}}}{=}g^{{\tt a}}\big[\mathscr{A}_{n-1}^{(m)}\big]+\sum_{\substack{n_L,m_L\\n_R,m_R}}g^{{\tt a}}\big[\mathscr{A}_{n_L}^{(m_L)}\big]\underset{\mathrm{BCFW}}{\bigotimes}g^{{\tt b}}\big[\mathscr{A}_{n_R}^{(m_R)}\big]\label{bcfw_in_momentum_twistors_deux},\vspace{-0.25cm}} where, as with the default contour prescription, {\it this same recursive rule is used for} every {\it lower-point amplitude}. This is illustrated in \mbox{Figure \ref{general_contour_figure}}.
\newpage
By varying the parameters {\tt \{a,b,c\}}, one can obtain a very wide-array of analytic formulae for any particular helicity amplitude. It could be that more general recursion-schemes will eventually prove useful,\footnote{For example, one could consider recursive schemes which make use of the parity-conjugate version of the BCFW-bridge, which make use of reflected (as well as rotated) lower-point amplitudes, or which allow for rotations of lower-point amplitudes to vary as a function of recursive depth. None of these generalizations are necessary for $n\leq9$, and we suspect that this is true generally.} but as far as we have been able to check, this class of recursion schemes has proven in some sense exhaustive. Specifically, we have checked that for up to 9-particles, this three-parameter family is sufficient to generate {\it all} linearly independent representations of superamplitudes. For example, there turn out to be 74 linearly-independent formulae for the 8-point N$^2$MHV tree amplitude, involving 176 Yangian-invariants. All of these formulae are worked-out explicitly as part of the demonstration file for the {\tt bcfw} package. 

There are three principle reasons why researchers may find this broad-class of tree-amplitude formulae useful. First, knowing the  range of possible tree-amplitude formulae helps one build intuition about amplitudes in general, and allows one to separate general properties about amplitudes from the peculiarities of particular formulae. Secondly, having many different representations available frees one from using unnecessarily inefficient representations of particular helicity amplitudes. For example, it is sometimes heard that ``the'' BCFW-formula (with the default scheme implicit) for the split-helicity amplitude is maximally-concise\footnote{This observation is true for the default recursion-scheme used by {\tt bcfw}; in particular, the helicity component $\mathscr{A}_{n}^{(m)}(-,\ldots,-,+\ldots,+)$ of {\tt generalTreeContour[0,0,0][n,m]} is the gluonic amplitude with the fewest number of non-vanishing BCFW terms; but this feature is observed for very few of the more general recursive schemes.} (meaning that a maximal number of terms in the tree-contour vanish); however, fixing a recursive scheme, this is true for at most {\it one particular} split-helicity amplitude---the other split-helicity amplitudes including some for which almost none of the BCFW terms vanish. And so, it should be possible to use the variety of representations that can be generated by {\tt bcfw} to find a `best-case' formula for any particular helicity amplitude of interest. And finally, because the BCFW formulae obtained using different recursive schemes often have very few spurious poles in common, it may be possible to combine a variety of BCFW formulae to avoid encountering spurious poles while generating Monte-Carlo events for phase-space integration, for example.

\begin{figure}[h]\begin{center}\caption{Examples of particular recursion schemes, highlighting how the lower-point amplitudes are rotated.\label{particular_schemes}}\vspace{0.1cm}
$\begin{array}{c}\text{{\tt bcfw}'s default scheme}\\[-0.5cm]\includegraphics[scale=0.5]{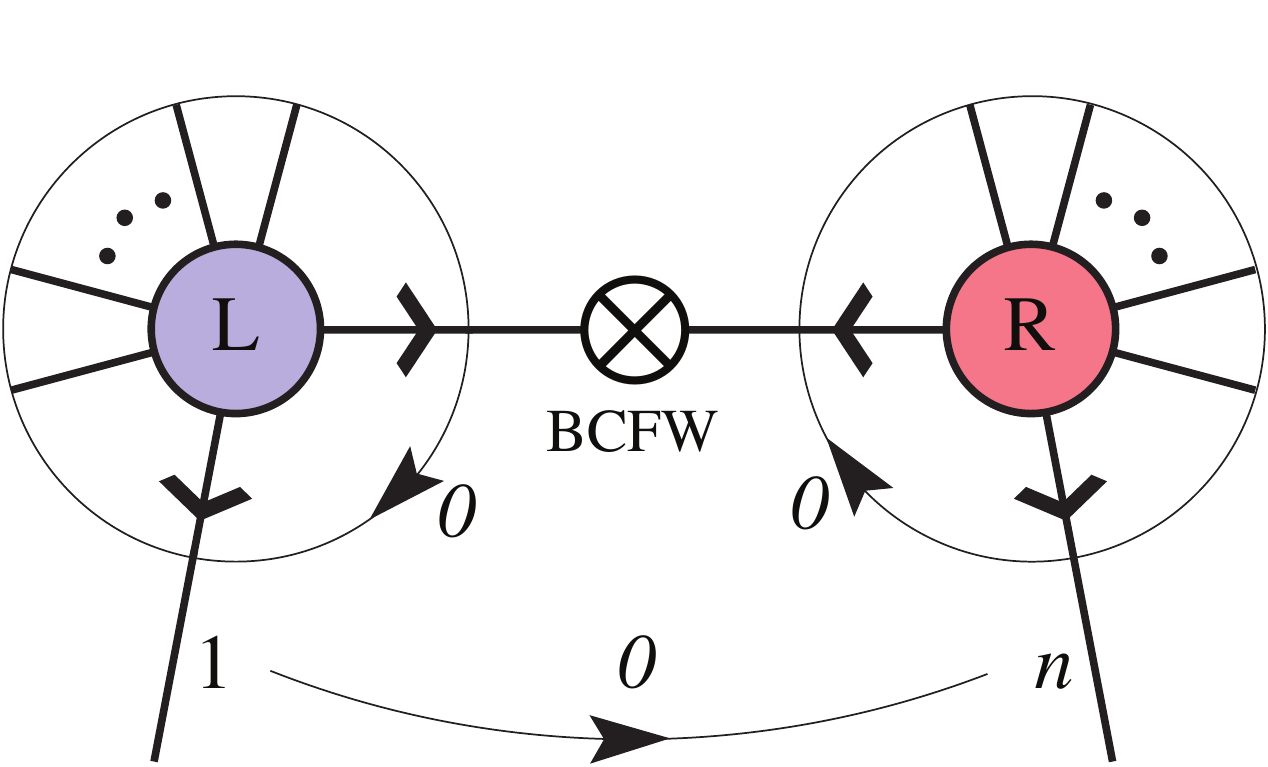}\\{\tt generalTreeContour[0,0,0]}\\\equiv{\tt treeContour}\end{array}$\hspace{1cm}$\begin{array}{c}\text{Drummond~\&~Henn's~scheme}\\[-0.5cm]\includegraphics[scale=0.5]{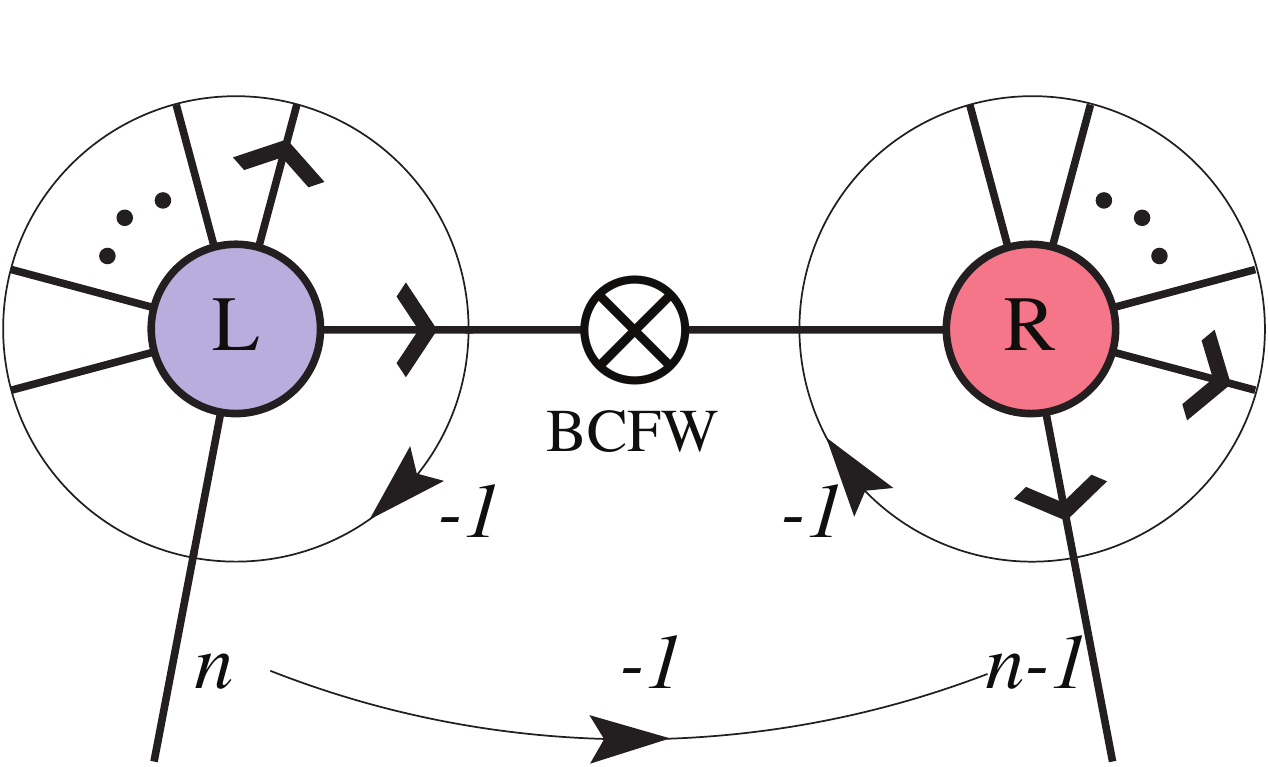}\\{\tt generalTreeContour[-1,-1,-1]}\\~\end{array}$\end{center}\vspace{-1cm}\end{figure}

It may be helpful to know that the particular recursive-scheme used by Drummond and Henn to solve the BCFW recursion relations in \cite{Drummond:2008cr}, corresponds to \linebreak {\tt generalTreeContour[-1,-1,-1]}; this scheme is illustrated in \mbox{Figure \ref{particular_schemes}}.

~\\

\subsection{Extracting Helicity Component-Amplitudes from Tree-Contours}
To compute a particular helicity amplitude from the supersymmetric contour integral, one need only project-out the desired Grassmann components, as dictated by the definition of the external superfields $\Phi_a^+$ given in equation (\ref{superfield}). Of course, the component fields of $\Phi_a^+$ are given in terms of $\tilde\eta$-variables, which, as described in \mbox{section \ref{kinematics_section}}, are related to the momentum-supertwistor Grassmann parameters $\eta_a$ via \eq{\tilde\eta_a=\qabinverse_{ab}\eta_b.} Because the matrix of coefficients of the Grassmann $\eta$'s for each {\tt residue} is nothing but its corresponding {\tt dMatrix}, we have that \eq{D_{\alpha\,a}\eta_{a}=D_{\alpha\,b}\qabinverse_{ba}\tilde\eta_a\equiv\hat{C}_{\alpha\,a}\tilde\eta_a.} 
In terms of the Grassmannian integral (\ref{grassmannian_contour}), this means that we may write
\begin{align}\hspace{-1.cm}({\tt residue})\prod_{\alpha=1}^{k}\delta^{0|4}\left(({\tt dMatrix})_{\alpha\,a}\eta_a\right)&=({\tt residue})\prod_{\alpha=1}^k\delta^{0|4}\left(({\tt dMatrix})_{\alpha\,b}({\tt QabInverse[n]})_{b\,a}\tilde{\eta}_a\right)\nonumber\\&\equiv({\tt residue})\prod_{\alpha=1}^k\delta^{0|4}\left(({\tt cHatMatrix})_{\alpha\,a}\tilde{\eta}_a\right).\end{align}
Upon explicitly including the full MHV super-amplitude we obtain,
\eqs{&\hspace{-0.9cm}\Longrightarrow \frac{({\tt residue})}{\ab{1\,2}\cdots\ab{n\,1}}\prod_{\underline{\alpha}=1}^{2}\delta^{0|4}\left(\lambda^{\underline{\alpha}}_{\,\,\,a}\tilde{\eta}_a\right)\prod_{\alpha=1}^k\delta^{0|4}\left(({\tt cHatMatrix})_{\alpha\,a}\tilde{\eta}_a\right)\\&\hspace{-0.9cm}\equiv\frac{({\tt residue})}{\ab{1\,2}\cdots\ab{n\,1}}\prod_{\hat{\alpha}=1}^{k+2}\delta^{0|4}\left({\tt cMatrix}_{\hat{\alpha}\,a}\tilde{\eta}_{a}\right),}
where we have defined the matrix $C_{\hat{\alpha}\,a}$ according to \eq{C_{\hat\alpha\,a}\equiv\left(\begin{array}{ccccccc}\hat{C}_{1\,1}&\hat{C}_{1\,2}&\cdots&\hat{C}_{1\,n-1}&\hat{C}_{1\,n}\\
\vdots&\vdots&\ddots&\vdots&\vdots\\
\hat{C}_{k\,1}&\hat{C}_{k\,2}&\cdots&\hat{C}_{k\,n-1}&\hat{C}_{k\,n}\\\hline\lambda_1^{\underline1}&\lambda_2^{\underline1}&\cdots&\lambda_{n-1}^{\underline1}&\lambda_n^{\underline1}\\\lambda_1^{\underline2}&\lambda_2^{\underline2}&\cdots&\lambda_{n-1}^{\underline2}&\lambda_n^{\underline2}\end{array}\right)\equiv\left(\begin{array}{c}\hat{C}_{\alpha\,a}\\\hline\lambda_a^{\underline\alpha}\end{array}\right).}
It is worth noting that just as each {\tt dMatrix} represents an isolated {\it point} in the Grassmannian of $k$-planes in $n$-dimensions, each {\tt cMatrix} gives an isolated {\it point} in the Grassmannian of $m(=k+2)$-planes in $n$-dimensions. Indeed, these are the isolated poles `encircled' by the (original) twistor-space Grassmannian contour-integral of \cite{ArkaniHamed:2009dn},
\eqs{\label{old_grassmannian_contour}\mathcal{A}_{n}^{(m=k+2)}&=\frac{1}{\mathrm{vol}(GL_m)}\!\oint\limits_{\Gamma_{n,m}}\!\!\frac{d^{n\times m} C_{\hat\alpha\,a}\,\prod_{\hat\alpha=1}^{m}\delta^{4|4}\left(C_{\hat\alpha\,a}\mathcal{W}_a\right)}{(1\cdots m)(2\cdots m\smallplus1)\cdots(n\cdots m\smallminus1)}.} The momentum-twistor Grassmannian integral (\ref{grassmannian_contour}) was derived from the original twistor-space integral (\ref{old_grassmannian_contour}) in \cite{ArkaniHamed:2009vw}, where it was shown how the MHV-prefactor arises naturally as {\it the Jacobian of the change-of-variables} in going from the (space-time) twistor-space variables $\mathcal{W}_a$ to the momentum-twistor-space variables $\mathcal{Z}_a$. 

Now, having the matrix of coefficients of the $\tilde\eta$-variables, it is particularly simple to extract any helicity component amplitude. For example, pure-glue amplitudes are given by 
\begin{align}\hspace{-0.5cm}\mathcal{A}_{n}^{(m)}(\ldots,j_1^{-},\ldots,j_{m}^-,\ldots)&=\int d^{0|4}\tilde\eta_{j_1}\cdots d^{0|4}\tilde\eta_{j_m}\Big[\mathcal{A}_{n}^{(m)}\Big];\\&=\sum_{\gamma\in\Gamma_{n,m}}\frac{({\tt residue}_{\gamma})}{\ab{1\,2}\cdots\ab{n\,1}}\left({\tt Det[cMatrix}_{\gamma}{\tt[[All,\{j_1,\ldots,j_m\}]]]}\right)^4.\nonumber\end{align}

More generally, each helicity amplitude can be `projected-out' of the superamplitude by multiplying each residue in the tree-contour by the appropriate set of four $(m\times m)$-minors of its corresponding matrix $C_{\hat{\alpha}\,a}$. The list of minors which project-out a particular helicity component-amplitude is given by the function {\tt parseInput[]}.

\section{The {\tt bcfw} {\sc Mathematica} Package}\label{bcfw_package_section}
A separate {\sc Mathematica} notebook---distributed along with {\tt bcfw.m}---has been prepared to introduce the reader to the many functions of {\tt bcfw} and their primary usage. We hope that the demonstration notebook is sufficiently self-contained for most users. In this section, we briefly describe the basic algorithmic structures which underly the {\tt bcfw} package, with an emphasis on the features that are likely to prove useful beyond the limited framework of {\sc Mathematica}. 

\subsection{Setup and Initialization}
Initialization of the package is simple: so long as the file being used has been saved to the same directory as the package's source {\tt bcfw.m}, one need only call the following:

\mathematica{1}{SetDirectory[NotebookDirectory[]];\hspace{8cm}<<bcfw.m}{\raisebox{-5.25cm}{\includegraphics[scale=0.75]{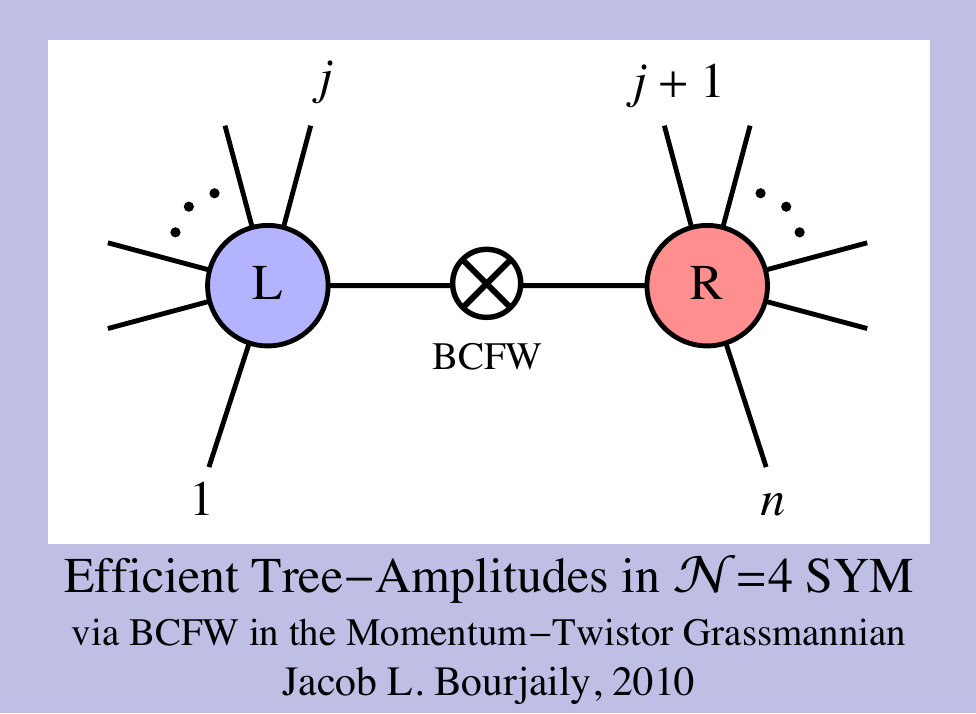}}}

\subsection{Getting Started with Analytic Tree Amplitudes}
To start gaining intuition for how helicity-amplitudes can be specified in {\tt bcfw}, consider a very simple example: the 8-point MHV amplitude $\mathcal{A}_8^{(2)}(+,+,-,+,+,-,+,+)$. This amplitude can easily be found using {\tt bcfw} through the command,

\mathematica{0.85}{Amp[p,p,m,p,p,m,p,p]}{$\displaystyle\frac{{\tt ab[3,6]}^{\tt 4}}{{\tt ab[1,2]ab[2,3]ab[3,4]ab[4,5]ab[5,6]ab[6,7]ab[7,8]ab[8,1]}}$}

To make the result more aesthetically appealing, any output of {\tt bcfw} can be wrapped by the function `{\tt nice[]}' which formats the result so that it is more ``human-readable.'' For example, using {\tt nice}, the above command would return:

\mathematica{0.65}{Amp[p,p,m,p,p,m,p,p]//nice}{$\displaystyle\frac{\ab{3\,6}^4}{\ab{1\,2}\ab{2\,3}\ab{3\,4}\ab{4\,5}\ab{5\,6}\ab{6\,7}\ab{7\,8}\ab{8\,1}}$}

\noindent We have chosen to make `{\tt nice}' formatting an `{\it opt-in}' option for users, so that the underlying structure is transparent at all times---and in order to avoid the pitfalls of conditional formatting in {\sc Mathematica} while maximizing the ease of symbolic manipulation.

Although the analytic formulae for tree amplitudes quickly become too long and complex for visual comprehension, {\tt bcfw}'s function {\tt Amp} will in fact write-out any amplitude. As one further example, consider the $6$-point NMHV alternating helicity amplitude. 

\vspace{-0.35cm}\mathematica{1}{Amp[m,p,m,p,m,p]//nice}{$\displaystyle\phantom{\,+\,} \frac{\ab{1\,5}^4 (\ab{3\,5} \ab{1\,2\,3\,4}-\ab{3\,4} \ab{1\,2\,3\,5})^4}{\ab{1\,2} \ab{2\,3} \ab{3\,4} \ab{4\,5} \ab{5\,6} \ab{6\,1} \ab{1\,2\,3\,4} \ab{1\,2\,3\,5} \ab{1\,2\,4\,5} \ab{1\,3\,4\,5} \ab{2\,3\,4\,5}}$\hspace{1cm}$\displaystyle-\frac{(\ab{1\,3} \ab{5\,6} \ab{1\,2\,3\,5}-\ab{1\,5} (\ab{3\,6} \ab{1\,2\,3\,5}+\ab{3\,5} \ab{2\,3\,6\,1}))^4}{\ab{1\,2} \ab{2\,3} \ab{3\,4} \ab{4\,5} \ab{5\,6} \ab{6\,1} \ab{1\,2\,3\,5} \ab{1\,2\,5\,6} \ab{1\,3\,5\,6} \ab{2\,3\,5\,6} \ab{2\,3\,6\,1}}$\hspace{1cm}$\displaystyle+\frac{(\ab{1\,3} \ab{5\,6} \ab{1\,3\,4\,5}-\ab{1\,5} (\ab{3\,6} \ab{1\,3\,4\,5}+\ab{3\,4} \ab{1\,3\,5\,6}+\ab{3\,5} \ab{3\,4\,6\,1}))^4}{\ab{1\,2} \ab{2\,3} \ab{3\,4} \ab{4\,5} \ab{5\,6} \ab{6\,1} \ab{1\,3\,4\,5} \ab{1\,3\,5\,6} \ab{3\,4\,5\,6} \ab{3\,4\,6\,1} \ab{4\,5\,6\,1}}$}

\vspace{-0.3cm}

We should emphasize, however, that direct evaluation of the formulae generated by {\tt Amp} (or {\tt AmpTerms}) are often {\it dramatically} less efficient than what can be obtained using {\tt nAmp} (or {\tt nAmpTerms}).\footnote{This is true even with fairly intelligent caching. Because of this, researchers interested in transferring the formulae generated by {\tt bcfw} to other frameworks should seriously consider using the {\it superamplitudes} directly.} This directly reflects the efficiency gained by the momentum-twistor Grassmanniannian representation of superamplitudes.\footnote{To better understand this, observe that each {\tt cMatrix} includes as its first $k$-rows the matrix {\tt cHatMatrix}$=${\tt dMatrix.QabInverse[n]}; this introduces many new kinematical invariants into each term---the two-brackets---while simultaneously duplicating each column of {\tt dMatrix} many times, greatly obfuscating an underlying simplicity with fundamentally redundant information.}

\begin{table}[t]\begin{center}\caption{6-point NMHV superamplitude $\mathcal{A}_6^{(3)}$, given by {\tt treeContour[6,3]}.\label{6_point_super_amp}}\mbox{\hspace{-2.05cm}\begin{tabular}{|@{$\,$}l@{$\,$}|@{$\,$}l@{$\,$}|@{$\,$}l@{$\,$}|@{$\,$}l@{$\,$}|}
\cline{2-4}
\multicolumn{1}{c@{$\,$}|@{$\,$}}{$~$}&\multicolumn{1}{@{$\,$}c@{$\,$}|@{$\,$}}{Name}&\multicolumn{1}{c@{$\,$}|@{$\,$}}{{\tt residue}}&\multicolumn{1}{c|}{{\tt dMatrix}}\\
\hline1.\mbox{\rule[40pt]{0pt}{-20pt}}&$R[1\,2\,3\,4\,5]$&$\displaystyle\frac{1}{\ab{1\,2\,3\,4}\ab{2\,3\,4\,5}\ab{3\,4\,5\,1}\ab{4\,5\,1\,2}\ab{5\,1\,2\,3}\ab{1\,2\,3\,4}}$&$\left(\begin{array}{cccccc}\ab{2\,3\,4\,5}&\ab{3\,4\,5\,1}&\ab{4\,5\,1\,2}&\ab{5\,1\,2\,3}&\ab{1\,2\,3\,4}&0\end{array}\right)$\\[10pt] 
\hline \mbox{\rule[40pt]{0pt}{-20pt}}2.&$R[1\,3\,4\,5\,6]$&$\displaystyle\frac{1}{\ab{1\,3\,4\,5}\ab{3\,4\,5\,6}\ab{4\,5\,6\,1}\ab{5\,6\,1\,3}\ab{6\,1\,3\,4}\ab{1\,3\,4\,5}}$&$\left(\begin{array}{cccccc}\ab{3\,4\,5\,6}&0&\ab{4\,5\,6\,1}&\ab{5\,6\,1\,3}&\ab{6\,1\,3\,4}&\ab{1\,3\,4\,5}\end{array}\right)$\\[10pt]
\hline3.\mbox{\rule[40pt]{0pt}{-20pt}}&$R[1\,2\,3\,5\,6]$&$\displaystyle\frac{1}{\ab{1\,2\,3\,5}\ab{2\,3\,5\,6}\ab{3\,5\,6\,1}\ab{5\,6\,1\,2}\ab{6\,1\,2\,3}\ab{1\,2\,3\,5}}$&$\left(\begin{array}{cccccc}\ab{2\,3\,5\,6}&\ab{3\,5\,6\,1}&\ab{5\,6\,1\,2}&0&\ab{6\,1\,2\,3}&\ab{1\,2\,3\,5}\end{array}\right)$
\\[10pt]\hline
\end{tabular}}\end{center}\end{table}
 
As described in the previous section, each superamplitude is represented by {\tt bcfw} as a contour integral in the momentum-twistor Grassmannian (\ref{grassmannian_contour}). The particular representation of the $n$-particle N$^{(m-2)}$MHV superamplitude derived via the BCFW recursion scheme with rotations {\tt\{a,b,c\}} is obtained with the function {\tt generalTreeContour[a,b,c][n,m]} (see section \ref{recursion_scheme_subsection}). The default representation---obtained using the default recursion scheme, with {\tt\{a,b,c\}=\{0,0,0\}}, is obtained with {\tt treeContour[n,m]}. For example, the default representation of the $6$-point NMHV superamplitude is given in \mbox{Table \ref{6_point_super_amp}}.

\subsection{Referencing, Generating, or Specifying Kinematical Data}
In order to evaluate amplitudes numerically using {\tt bcfw}, kinematical data must first be defined. This can be done by calling upon a list of reference momentum-twistors, freshly-generating random kinematics, or by specifying kinematical data explicitly:
\begin{enumerate}
\item{\bf {\tt useReferences[n]:}} use a standard set of reference momentum-twistors; these reference twistors were carefully selected so that \begin{itemize}
\item all components are integer-valued (and small);
\item there are no physical or spurious singularities;
\item all kinematical invariants are uniformly {\it positive} (that is, $s_{a\ldots b}>0$ for all ranges $a\ldots b$), and that these invariants are numerically given by ratios of relatively small integers---leading to amplitudes that are ratios of integers that are `not-too-horrendously-long';
\end{itemize}
\begin{table}[h!]\begin{center}\caption{Reference momentum-twistors used in {\tt bcfw}'s function {\tt useReferences[n].}\label{reference_momentum_twistors}}\vspace{-0.4cm}\mbox{\hspace{-0.35cm}\begin{tabular}{|c|r|r|r|r|r|r|r|r|r|r|r|r|r|r|r|}\cline{2-16}\multicolumn{1}{c|}{$~$}&$Z_{1}$&$Z_{2}$&$Z_{3}$&$Z_{4}$&$Z_{5}$&$Z_{6}$&$Z_{7}$&$Z_{8}$&$Z_{9}$&$Z_{10}$&$Z_{11}$&$Z_{12}$&$Z_{13}$&$Z_{14}$&$Z_{15}$\\\hline$Z_a^{1}$&$-3$&$2$&$-2$&$3$&$0$&$-1$&$2$&$2$&$4$&$-2$&$-5$&$-1$&$5$&$6$&$4$\\$Z_a^{2}$&$5$&$6$&$5$&$3$&$-5$&$2$&$0$&$1$&$-1$&$-5$&$2$&$6$&$-5$&$4$&$6$\\$Z_a^{3}$&$3$&$-1$&$-1$&$5$&$6$&$-5$&$-6$&$-5$&$-6$&$4$&$6$&$1$&$-5$&$-5$&$-3$\\$Z_a^{4}$&$-3$&$-3$&$5$&$-2$&$0$&$-5$&$-1$&$-3$&$1$&$4$&$-1$&$-4$&$3$&$-3$&$-4$\\\hline\end{tabular}}
\end{center}\vspace{-0.8cm}\end{table}
\begin{figure}[h!]\begin{center}\caption{Evaluation of 10-point N$^3$MHV helicity amplitudes to infinite precision using {\it reference momentum-twistors}. The timing reflects the fact that the first computation determined the full superamplitude {\it and} projected-out a particular helicity component, while the second only needed to perform the projection.\label{reference_10_pt}}\vspace{-0.3cm}\mbox{\hspace{-0.04\textwidth}\vspace{0.35cm}\noindent\boxed{\begin{minipage}{1.07\textwidth}\begin{tabular}{lp{11cm}}\\[0.00cm]
{\color{paper_blue}{\scriptsize{\tt In[1]:=}}}&\vspace{-1.00cm}{\tt useReferences[10];\linebreak~nAmp[m,m,m,m,m,p,p,p,p,p]//withTiming}\\[0.9cm]
{\color{paper_blue}{\scriptsize {\tt Out[1]:=}}}&\vspace{-1.25cm}{\footnotesize \mbox{\qquad{\tt Evaluation of the 10-point N$^3$MHV amplitude required {\color{paper_blue}46.7.\ ms} to complete.}}}\vspace{0.3cm}\linebreak~\mbox{{\small$\hspace{-0.25cm}\displaystyle\frac{{\tt 17886892256634020134576330754470391777}}{{\tt 280278666971743564282064966167680000}}$}}\\[0.2cm]
{\color{paper_blue}{\scriptsize{\tt In[2]:=}}}&{\tt nAmp[m,p,m,p,m,p,m,p,m,p]//withTiming}\\[0.9cm]
{\color{paper_blue}{\scriptsize {\tt Out[2]:=}}}&\vspace{-1.25cm}{\footnotesize \mbox{\qquad{\tt Evaluation of the 10-point N$^3$MHV amplitude required {\color{paper_blue}8.6.\ ms} to complete.}}}\vspace{0.3cm}\linebreak~\mbox{{\small$\hspace{-0.25cm}-\displaystyle\frac{{\tt 5007045380847632725336670465304701314367799201604575059832902148541}}{{\tt 213450466354689126392301641566350924968168379805192061706240000}}$}}
\end{tabular}\end{minipage}}\vspace{0.35cm}}
\end{center}\vspace{-0.5cm}\end{figure}

$\qquad$In \mbox{Table \ref{reference_momentum_twistors}} we give a sample of the reference momentum-twistors. Because an arbitrary set of momentum-twistors define on-shell, momentum-conserving kinematics, there are no constraints from momentum conservation. Therefore, choosing simply the first $n$ twistors from the list in \mbox{Table \ref{reference_momentum_twistors}} will suffice. It is worth mentioning, however, that these reference momentum-twistors are neither canonically normalized\footnote{By not having canonical normalization, we mean that there are non-trivial, Lorentz-frame (and hence also little-group)-dependent kinematical scale-factors in the spinors; however, this tends to only cause a problem when combining/comparing multiple helicity component-amplitudes.}, nor do they map to {\it real} four-momenta in \nolinebreak$\mathbb{R}^{3,1}$. 

$\qquad$ Nonetheless, reference twistors are extremely well-suited for debugging, checking identities, and finding relations to infinite precision. As one can see in \mbox{Figure \ref{reference_10_pt}}, using {\tt bcfw}'s built-in reference momentum-twistors can quickly lead to scattering amplitudes that are known to infinite-precision. Notice that in \mbox{Figure \ref{reference_10_pt}}, once the {\it super}amplitude has been computed for any helicity-component, all subsequent components are obtained quite rapidly. 

\item {\bf{\tt useRandomKinematics[n]:}} use randomly-generated kinematics in $\mathbb{R}^{3,1}$. This function chooses a random set of (optionally rational or arbitrary-precision) on-shell four-momenta in $\mathbb{R}^{3,1}$, and sets up essentially all the kinematical variables of potential interest, including 
\begin{itemize}
\item momentum-twistors $\{\overrightarrow{Z}\}\equiv${\tt Zs}, given as an $(n\times 4)$ matrix---the $n$ rows listing the four homogeneous components of each momentum-twistor; \linebreak {\tt useRandomKinematics[n]} also defines the `{\it dual}' momentum-twistors \linebreak $\{\overrightarrow{W}\}\equiv${\tt Ws}, which, although not used by {\tt bcfw}, may be found useful by some researchers;
\item spinor-helicity variables $\{\overrightarrow\lambda\}\equiv${\tt Ls} and $\{\overrightarrow{\tilde\lambda}\}\equiv${\tt Lbs}, each an $(n\times 2)$ matrix of components; these have been normalized so that $\tilde\lambda_a=\pm\left(\lambda_a\right)^*$, as described in \mbox{section \ref{kinematics_section}};
\item {\tt fourMomenta}, an $(n\times4)$ matrix of the components $(p^0,p^x,p^y,p^z)$ of each four-momentum;
\item {\tt regionMomenta}, the dual-coordinates (described in \mbox{section \ref{kinematics_section}}), given as a $n$-length list of $2\times2$ Hermitian matrices;
\end{itemize}
\begin{figure}[t!]\begin{center}\caption{Evaluation of 12-point N$^4$MHV helicity amplitudes with random kinematics. \label{reference_12_pt}}\vspace{-0.3cm}\mbox{\hspace{-0.04\textwidth}\vspace{0.35cm}\noindent\boxed{\begin{minipage}{1.07\textwidth}\begin{tabular}{lp{11cm}}\\[0.20cm]
{\color{paper_blue}{\scriptsize{\tt In[1]:=}}}&\vspace{-1.00cm}{\tt useRandomKinematics[12];\linebreak~nAmp[m,p,m,p,m,p,m,p,m,p,m,p]//withTiming}\\[0.5cm]
{\color{paper_blue}{\scriptsize {\tt Out[1]:=}}}&\vspace{-0.85cm}{\footnotesize \mbox{\qquad{\tt Evaluation of the 12-point N$^4$MHV amplitude required {\color{paper_blue}596.\ ms} to complete.}}}\vspace{-0.1cm}\linebreak~\mbox{{\small$\hspace{-0.25cm}\displaystyle{\tt-274.127-5171.81~I}$}}\\[0.0cm]
{\color{paper_blue}{\scriptsize{\tt In[2]:=}}}&{\tt nAmp[p,m,p,m,p,m,p,m,p,m,p,m]//withTiming}\\[0.5cm]
{\color{paper_blue}{\scriptsize {\tt Out[2]:=}}}&\vspace{-0.85cm}{\footnotesize \mbox{\qquad{\tt Evaluation of the 12-point N$^4$MHV amplitude required {\color{paper_blue}71.4.\ ms} to complete.}}}\vspace{-0.1cm}\linebreak~\mbox{{\small$\hspace{-0.25cm}\displaystyle{\tt-274.127+5171.81~I}$}}
\end{tabular}\end{minipage}}\vspace{0.35cm}}
\end{center}\vspace{-0.3cm}\end{figure}
$\qquad$ An example of using random kinematics is shown in \mbox{Figure \ref{reference_12_pt}}, where the two alternating-helicity 12-point N$^4$MHV amplitudes were evaluated. Notice as before that once the superamplitude has been evaluated, subsequent helicity components are quickly extracted. Also, observe that the randomly-generated spinors and momentum-twistors have been appropriately normalized so that parity-conjugation results in complex-conjugation of the amplitude. 
~\\
\item using user-defined kinematics, given in terms of:
\begin{enumerate}
\item{\bf {\tt setupUsingFourMomenta[fourMomentaList]:}} generates momentum-twistors and spinor-helicity variables for the input list of four-momenta, \linebreak{\tt fourMomentaList}, which must be given as an $n$-tuple of four-vectors listing the components of each four-momentum; the list of four-momenta must conserve momentum;
\item{\bf {\tt setupUsingSpinors[Ls,Lbs]:}} generates momentum-twistors given the spinor-helicity variables {\tt Ls}$\equiv\{\overrightarrow{\lambda}\}$ and {\tt Lbs}$\equiv\{\overrightarrow{\tilde\lambda}\}$ each given as an $(n\times 2)$ matrix of components;
\item{\bf {\tt setupUsingTwistors[twistorList]:}} establishes the necessary kinematical functions given the (unconstrained) list of user-generated momentum-twistors. 
\end{enumerate}
$\qquad$ Examples of how each of these functions can be used can be found in the demonstration file included with the {\tt bcfw} package. 
\end{enumerate}

\subsection{Numerical Evaluation of Tree Amplitudes}
As has been emphasized throughout this paper, the principle sources of {\tt bcfw}'s efficiency are manifest supersymmetry and the use of momentum-twistor variables, which are both made manifest in the momentum-twistor Grassmannian integral (\ref{grassmannian_contour}). Because these ingredients---or at least their implementation---are quite novel in {\tt bcfw}, it is worth describing in some detail how amplitudes are evaluated numerically by the {\tt bcfw} package. 

The basic evaluation strategy is outlined in \mbox{Table \ref{detailed_timing}}, where we give the basic evaluation times for each step in the evaluation of the 10-point N$^3$MHV alternating-helicity tree-amplitude. 

Because of the central role played by momentum-twistors, the first step of any numerical evaluation is the establishment of momentum-twistor variables which can then be used to compute the kinematical invariants that determine any scattering amplitude. This can be done in a number of different ways---as described in the previous subsection. Although this should be completely clear from the discussions above, this step is not very computationally-intensive (and indeed, can be discounted entirely by choosing to randomly-generate momentum-twistors instead of four-momenta).

Because of the ubiquity of the MHV-amplitude pre-factor, $1/(\ab{1\,2}\cdots\ab{n\,1})$, and the map $\qabinverse_{ab}$ used to relate the momentum-twistors' $\eta$-variables to the $\tilde\eta$ variables of the external superfields, {\tt bcfw} evaluates these two objects and stores them globally whenever new kinematical data is defined. 

The first step in the evaluation of any particular helicity amplitude is actually the evaluation of the {\it full superamplitude}---represented as the list of BCFW-terms, where each is described by the pair {\tt \{residue, dMatrix\}} (which is stored in memory as the function {\tt nContour[a,b,c][n,m]}). Because particular helicity amplitudes are usually specified with respect to the $\tilde\eta$-variables of the external superfields, the {\tt dMatrix} of each {\tt residue} is then converted to the corresponding {\tt cMatrix} as described in section \ref{kinematics_section}. 

Once each BCFW-term has been evaluated numerically and stored as the pair {\tt \{residue,cMatrix\}}, it is relatively easy to extract any particular helicity component amplitude---by multiplying each {\tt residue} by the appropriate four $(m\times m)$ minors of its corresponding {\tt cMatrix}. This last step is nothing exotic: it is merely the evaluation of the Grassmann integrals $\int \prod_{i=1}^{m}d^{0|4}\tilde\eta_i$ which project-out a helicity-component amplitude from the superamplitude.

\begin{table}[t!]\vspace{-0.4cm}\begin{center}\caption{The general evaluation strategy used by {\tt bcfw}, with a break-down of evaluation-time requirements for each step in the case of the alternating-helicity 10-point N$^3$MHV tree-amplitude, $\mathcal{A}_{10}^{(5)}\big(-,+,-,+,-,+,-,+,-,+\big)$ (for random kinematics).\label{detailed_timing}}\boxed{\begin{minipage}{\textwidth}\begin{minipage}{0.87\textwidth}\begin{enumerate}
\item {\tt setupUsingRandomKinematics[10]}\vspace{-0.3cm}
\begin{enumerate}
\item generate random (on-shell, rational, momentum-conserving) four-momenta in $\mathbb{R}^{3,1}$; define spinors and momentum-twistors
\end{enumerate}\end{enumerate}\end{minipage}

\vspace{-12.0pt}\begin{minipage}{\textwidth}\hspace{\fill}{\bf 3.61 ms}\end{minipage}

\noindent\begin{minipage}{0.87\textwidth}\begin{enumerate}\item[]\begin{enumerate}\addtocounter{enumii}{1}
\item evaluate the universal objects {\tt nMHVprefactor} and {\tt nQinverse}
\end{enumerate}\end{enumerate}\end{minipage}

\vspace{-17.pt}\begin{minipage}{\textwidth}\hspace{\fill}{\bf 1.34 ms}\end{minipage}

\vspace{0.2cm}\noindent\begin{minipage}{0.87\textwidth}\begin{enumerate}\addtocounter{enumi}{1}
\item {\tt nAmp[m,p,m,p,m,p,m,p,m,p]}\vspace{-0.3cm}
\begin{enumerate}
\item evaluate the full-superamplitude, which is stored stored as the function {\tt nContour[0,0,0][10,5]} (for possible future use)
\end{enumerate}\end{enumerate}\end{minipage}

\vspace{-12pt}\begin{minipage}{\textwidth}\hspace{\fill}{\bf 23.2 ms}\end{minipage}

\noindent\begin{minipage}{0.9\textwidth}\begin{enumerate}\item[]\begin{enumerate}\addtocounter{enumii}{1}
\item convert each {\tt dMatrix} to its corresponding {\tt cMatrix}\end{enumerate}\end{enumerate}\end{minipage}

\vspace{-17pt}\begin{minipage}{\textwidth}\hspace{\fill}{\bf 3.03 ms}\end{minipage}

\noindent\begin{minipage}{0.9\textwidth}\begin{enumerate}\item[]\begin{enumerate}\addtocounter{enumii}{2}
\item project-out the desired helicity component-amplitude\end{enumerate}\end{enumerate}\end{minipage}

\vspace{-17pt}\begin{minipage}{\textwidth}\hspace{\fill}{\bf 4.01 ms}\end{minipage}
~\\
\begin{minipage}{\textwidth}\hspace{-0.0cm}{\bf Total Time:}\hspace{\fill}{\bf 35.2 ms}\end{minipage}\end{minipage}}\end{center}\end{table}

\newpage

\subsection{Example Applications}
In the demonstration file which accompanies the {\tt bcfw} package, several examples are given which illustrate how {\tt bcfw} can be used as a tool to verify results, find identities, or learn about scattering amplitudes more generally. In particular, these examples emphasize how using integer-valued reference momentum-twistors to compute amplitudes (and individual BCFW-terms) to infinite-precision can prove quite useful theoretically. The examples include:
\begin{itemize}
\item a verification of supersymmetric Ward identities; in particular, we check one of the `cyclic' identities described in \cite{Elvang:2009wd} for the 10-point N$^3$MHV amplitude---
\eqs{\hspace{-1cm}0=&\phantom{\,+\,}\mathcal{A}_{10}^{(5)}\left(\psi_{-1/2}^{(123)},\psi_{+1/2}^{\mathbf{(3)}},\psi_{+1/2}^{(1)},\psi_{+1/2}^{\mathbf{(4)}},\psi_{+1/2}^{(3)},\phi_{0}^{(2\mathbf{4)}},\phi_{0}^{(1\mathbf{4})},\phi_0^{(12)},\psi_{-1/2}^{(234)},g_{-}^{(1234)}\right)\\
&+\mathcal{A}_{10}^{(5)}\left(\psi_{-1/2}^{(123)},\psi_{+1/2}^{\mathbf{(4)}},\psi_{+1/2}^{(1)},\psi_{+1/2}^{\mathbf{(4)}},\psi_{+1/2}^{(3)},\phi_0^{(2\mathbf{4})},\phi_0^{(1\mathbf{3})},\phi_0^{(12)},\psi_{-1/2}^{(234)},g_-^{(1234)}\right)\\
&+\mathcal{A}_{10}^{(5)}\left(\psi_{-1/2}^{(123)},\psi_{+1/2}^{\mathbf{(4)}},\psi_{+1/2}^{(1)},\psi_{+1/2}^{\mathbf{(4)}},\psi_{+1/2}^{(3)},\phi_0^{(2\mathbf{3})},\phi_0^{(1\mathbf{4})},\phi_0^{(12)},\psi_{-1/2}^{(234)},g_-^{(1234)}\right)\\
&+\mathcal{A}_{10}^{(5)}\left(\psi_{-1/2}^{(123)},\psi_{+1/2}^{\mathbf{(4)}},\psi_{+1/2}^{(1)},\psi_{+1/2}^{\mathbf{(3)}},\psi_{+1/2}^{(3)},\phi_0^{(2\mathbf{4})},\phi_0^{(1\mathbf{4})},\phi_0^{(12)},\psi_{-1/2}^{(234)},g_-^{(1234)}\right)\label{susy_ward_identity}}---the verification of which is illustrated in \mbox{Figure \ref{mathematica_ward_identity}}, highlighting the power of knowing amplitudes to infinite precision;  
\item an explicit verification of the $U_1$-decoupling identity for the 10-point N$^3$MHV tree-amplitude (which, although a trivial consequence of any Lagrangian field theory, is a highly non-trivial check of numerical code!\footnote{We thank Freddy Cachazo for this suggestion.});
\item a complete classification of the linearly-independent BCFW-generated formulae for the 8-point N$^2$MHV supersymmetric tree-amplitude.
\end{itemize}


\begin{figure}[h]\begin{center}\caption{Verifying a supersymmetric Ward identity of the 10-point N$^3$MHV amplitude\label{mathematica_ward_identity}.}{\vspace{-0.35cm}\noindent\boxed{\begin{minipage}{1\textwidth}\begin{tabular}{lp{11cm}}\\[0.1cm]
{\color{paper_blue}{\scriptsize{\tt In[1]:=}}}&\vspace{-1.05cm}{\tt useReferences[10]}\linebreak~\vspace{-0.225cm}{\footnotesize{\tt \eqs{\hspace{-0.00cm}{\tt List\Big[}&
{\tt nAmp[\{1,2,3\},\{\mathbf{3}\},\{1\},\{\mathbf{4}\},\{3\},\{2,\mathbf{4}\},\{1,\mathbf{4}\},\{1,2\},\{2,3,4\},\{1,2,3,4\}]}\\&
{\tt nAmp[\{1,2,3\},\{\mathbf{4}\},\{1\},\{\mathbf{4}\},\{3\},\{2,\mathbf{4}\},\{1,\mathbf{3}\},\{1,2\},\{2,3,4\},\{1,2,3,4\}],}\\&
{\tt nAmp[\{1,2,3\},\{\mathbf{4}\},\{1\},\{\mathbf{4}\},\{3\},\{2,\mathbf{3}\},\{1,\mathbf{4}\},\{1,2\},\{2,3,4\},\{1,2,3,4\}],}\\&
{\tt nAmp[\{1,2,3\},\{\mathbf{4}\},\{1\},\{\mathbf{3}\},\{3\},\{2,\mathbf{4}\},\{1,\mathbf{4}\},\{1,2\},\{2,3,4\},\{1,2,3,4\}]\Big]} \nonumber}}}\\[-0.2cm]
{\color{paper_blue}{\scriptsize {\tt Out[1]:=}}}&\vspace{-1.475cm}{\small{\tt\eqs{\hspace{-3.8cm}\Bigg\{&\phantom{-\,}\mathtt{\frac{79370862801471295255}{28753113503920775424},}\hspace{7.5cm}~\\&\phantom{-\,}\mathtt{\frac{1275513453387873135869428633786428491}{77923676342112832490222204964602880},}\\&\mathtt{-\frac{40428898488502522106856665437052838463}{10951273590541549612279689882333035520},}\\&\mathtt{-\frac{16319258699414773847825256760953737}{1057119835135513498965174929610240}}\Bigg\}\nonumber}}}\\[-0.6cm]
{\color{paper_blue}{\scriptsize{\tt In[2]:=}}}&\hspace{-0.0cm}{\tt $\mathtt{Total[Out[1]]}\hspace{-0.55cm}$}\\[0.1cm]
\\[-0.6cm]{\color{paper_blue}{\scriptsize {\tt Out[2]:=}}}&{\tt 0}
\end{tabular}\end{minipage}}\vspace{-0.35cm}}\end{center}\end{figure}

~\newpage

\section{Comparison of {\tt bcfw} with Other Computational Tools}\label{comparison_section}
Although we do hope that the {\tt bcfw} package will prove useful in computations directly relevant to collider physics, its primary role will likely be as a tool for gaining intuition about scattering amplitudes, checking results/conjectures, and as a working example of a novel computational strategy that could perhaps be implemented much more efficiently by researchers with more computational expertise by optimizing either hardware or software.

Although there exists a wide variety of tools for computing scattering amplitudes---including {\tt COMIX} \cite{Gleisberg:2008fv}, {\tt AMEGIC++} \cite{Krauss:2001iv}, {\tt CompHEP} \cite{Pukhov:1999gg}, {\tt MadGraph} \cite{Maltoni:2002qb}, {\tt HELAC} \cite{Kanaki:2000ey}, and {\tt ALPHA} \cite{Caravaglios:1995cd}---it would be difficult for us to make any just comparison between these packages and {\tt bcfw}. (But it would be very interesting to see how the representation of amplitudes used by {\tt bcfw} would compare with the results of \cite{Dinsdale:2006sq}, or the impressive algorithms described in \cite{Giele:2010ks} based on the Berends-Giele recursion relations \cite{Berends:1987me}.) This is both because of the inherent inefficiencies of any {\sc Mathematica} package relative to compiled code, and also because these packages make use of a wide variety of different computational strategies and have widely-different scopes of purpose. 

Nevertheless, at least within the limited scope of {\sc Mathematica}, the recent release of the public package {\tt Gluon-Gluiono-Trees} ({\tt GGT}) \cite{Dixon:2010ik} has made at least a passing comparison between the two packages justified. After all, both packages were written in and for {\sc Mathematica}, both are based on the BCFW recursion relations, and both are (in principle) capable of computing all pure-glue amplitudes in $\mathcal{N}=4$, giving ample room for comparison. Moreover, because {\tt GGT} uses the form of tree-amplitudes in $\mathcal{N}=4$ obtained by Drummond and Henn in \cite{Drummond:2008cr}---which corresponds to the recursion scheme with {\tt\{a,b,c\}=\{-1,-1,-1\}} (see section \ref{recursion_scheme_subsection}), the amplitudes computed in {\tt GGT} match {\it term-by-term} the output of {\tt nAmpTerms[-1,-1,-1][helicityConfiguration]}, it is natural to wonder how the two packages compare in efficiency (at least in gross terms). 

The two packages {\tt GGT} and {\tt bcfw} were compared in passing in section \ref{introduction}, where we listed in \mbox{Table \ref{unpolarized_times}}, the times required to compute {\it unpolarized} (colour-ordered) $n$-gluon scattering cross sections. Considering that both packages were written for largely-theoretical purposes (as demonstrations of algorithms, for intuition-building, and checking results), it is easy to argue that the computation of all 1002 helicity-amplitudes for 10 gluons is not the computation either package was designed for.

A more reasonable comparison, and one which may shed light on the source of the disparity between the two packages, would be the evaluation of an individual helicity amplitudes. Consider for example the 8-particle N$^2$MHV alternating-helicity amplitude. This computation is illustrated in \mbox{Figure \ref{8point_comparison}}. Here, we have explicitly separated the time required by {\tt GGT} to convert the analytic amplitude to a form suitable for numerical evaluation and the time actually required for evaluation by {\tt S@M}.\footnote{As indicated in \mbox{Figure \ref{8point_comparison}}, over half of the computation time is absorbed by the third-party package used by {\tt GGT}, `{\tt S@M}' \cite{Maitre:2007jq}, to evaluate formulae written in its spinor-helicity formalism. Although this package is far from optimal, it seems unlikely to account for more than a small fraction of the disparity between the two packages.} Of course, only the latter time is of essential interest in computations, as this represents (perhaps poorly) the essential complexity of the analytic forms of the amplitudes generated by the two packages' frameworks. 
(We should mention in passing that \mbox{Figure \ref{8point_comparison}} is not really a fair comparison between the two packages: in its \mbox{$4$ milliseconds}, {\tt bcfw} actually evaluated the entire 8-particle N$^2$MHV {\it superamplitude}---only a half-millisecond of which was used to project-out the alternating-helicity component-amplitude.)
\begin{figure}[t!]\vspace{-0.2cm}\begin{center}\caption{Comparing the evaluation of $\mathcal{A}_8^{(4)}(+,-,+,-,+,-,+,-)$ in {\tt GGT/S@M} vs.\ {\tt bcfw}.\label{8point_comparison}}\vspace{-0.35cm}\mbox{\hspace{-0.025\textwidth}\noindent\boxed{\begin{minipage}{1.05\textwidth}\begin{tabular}{lp{11cm}}
{\color{paper_blue}{\scriptsize{\tt In[1]:=}}}&{\tt spinorForm=GGTtoSpinors[GGTnnmhvgluon[8,2,4,6]]//withTiming;}\\
{\color{paper_blue}{\scriptsize{\tt In[2]:=}}}&{\tt N[spinorForm]//withTiming}\\
{\color{paper_blue}{\scriptsize {\tt ~}}}&{\footnotesize \mbox{{\tt$~\;$Evaluation of the function $GGTtoSpinors$ required {\color{paper_blue}4.29 seconds} to complete.}}}\\
{\color{paper_blue}{\scriptsize {\tt ~}}}&{\footnotesize \mbox{{\tt$~\;$Evaluation of the function $N$ required {\color{paper_blue}5.84 seconds} to complete.}}}\\
{\color{paper_blue}{\scriptsize {\tt Out[2]:=}}}&{\tt -0.395021 \,+\, 0.310719\,I}\\
{\color{paper_blue}{\scriptsize{\tt In[3]:=}}}&{\tt nAmp[p,m,p,m,p,m,p,m]//withTiming}\\
{\color{paper_blue}{\scriptsize {\tt ~}}}&{\footnotesize \mbox{{\tt $~\;$Evaluation of the 8-point N$^2$MHV amplitude required {\color{paper_blue}4.30 ms} to complete.}}}\\
{\color{paper_blue}{\scriptsize {\tt ~}}}&{\footnotesize \mbox{{\tt $~\;\;\;\;$$\mathcal{A}_{8}^{(4)}(g_+,g_-,g_+,g_-,g_+,g_-,g_+,g_-):$}}}\\
{\color{paper_blue}{\scriptsize {\tt Out[3]:=}}}&{\tt -0.395021 \,+\, 0.310719\,I}\\
\end{tabular}\end{minipage}}}\vspace{-0.35cm}\end{center}\vspace{-0.55cm}\end{figure}

A better understanding the relative efficiency between the two packages requires a more systematic survey than the isolated example \mbox{Figure \ref{8point_comparison}}. To give an idea of how the two packages compare more generally, \mbox{Table \ref{particular_evaluation_times}} lists the kinematically-averaged evaluation times for a range of particular pure-glue scattering amplitudes. Similar time comparisons could have been made for amplitudes involving gluinos (it is worth remembering that because {\tt bcfw} always computes the entire superamplitude before extracting a component, the times quoted in \mbox{Table \ref{particular_evaluation_times}} would be roughly the same regardless of the helicity components used for comparison).

The scale of the differences observed in \mbox{Table \ref{particular_evaluation_times}} is hard to overlook, and naturally raises the question of what underlies the difference in efficiency? This question seems especially relevant considering that if the particular BCFW recursion scheme {\tt nAmpTerms[-1,-1,-1]} is used to compute amplitudes in {\tt bcfw} the two packages agree term-by-term for every component amplitude (including those involving gluinos). We suspect that the two main sources of relative efficiency between {\tt bcfw} and {\tt GGT} are simply: the use of momentum-twistor variables, and keeping supersymmetry manifest throughout every computation, by describing amplitudes directly as contour integrals in the Grassmannian. 

\begin{table}[t!]\begin{center}\caption{\mbox{Comparing evaluation-times in {\tt bcfw} and {\tt GGT/S@M} for particular helicity amplitudes.\label{particular_evaluation_times}}}
\vspace{-0.3cm}\mbox{\hspace{-2.05cm}\begin{tabular}{|l|l|l|}
\cline{2-3}\multicolumn{1}{c|}{$$}&\multicolumn{2}{l|}{{\bf Time ($10^{-3}$ s)}}\\\hline
\multicolumn{1}{|c|}{{\bf NMHV Amplitudes}}& {\tt bcfw}&\multicolumn{1}{c|}{{\tt GGT}}\\\hline
$\mathcal{A}_{5}^{(3)}(-,+,-,+,-)$&$0.67$&$16$\\
$\mathcal{A}_{6}^{(3)}(-,+,-,+,-,+)$&$0.98 $&$71 $\\
$\mathcal{A}_{7}^{(3)}(-,+,-,+,-,+,+)$&$1.4 $&$240 $\\
$\mathcal{A}_{8}^{(3)}(-,+,-,+,-,+,+,+)$&$1.9$&$690$\\
$\mathcal{A}_{9}^{(3)}(-,+,-,+,-,+,\cdots,+)$&$2.6$&$1,\!100$\\
$\mathcal{A}_{10}^{(3)}(-,+,-,+,-,+,\cdots,+)$&$3.5$&$2,\!000$\\
$\mathcal{A}_{11}^{(3)}(-,+,-,+,-,+,\cdots,+)$&$4.6$&$3,\!400$\\
$\mathcal{A}_{12}^{(3)}(-,+,-,+,-,+,\cdots,+)$&$5.7$&$5,\!000$\\\hline
\end{tabular}}\;\raisebox{-0.30cm}{\begin{tabular}{|l|l|l|}
\cline{2-3}\multicolumn{1}{c|}{$$}&\multicolumn{2}{l|}{{\bf Time ($10^{-3}$ s)}}\\\hline
\multicolumn{1}{|c|}{{\bf N}$^{\mathbf{2}}${\bf MHV Amplitudes}}& {\tt bcfw}&\multicolumn{1}{c|}{{\tt GGT}}\\\hline
$\mathcal{A}_{6}^{(4)}(-,+,-,+,-,-)$&$0.91$&$230$\\
$\mathcal{A}_{7}^{(4)}(-,+,-,+,-,+,-)$&$2.0 $&$2,\!300$\\
$\mathcal{A}_{8}^{(4)}(-,+,-,+,-,+,-,+)$&$4.2$&$16,\!000$\\
$\mathcal{A}_{9}^{(4)}(-,+,-,+,-,+,-,+,+)$&$8.6$&$72,\!000$\\
$\mathcal{A}_{10}^{(4)}(-,+,-,+,-,+,-,+,+,+)$&$16 $&$260,\!000$\\
$\mathcal{A}_{11}^{(4)}(-,+,-,+,-,+,-,+,\cdots,+)$&$30 $&$740,\!000$\\
$\mathcal{A}_{12}^{(4)}(-,+,-,+,-,+,-,+,\cdots,+)$&$48 $&$1,\!900,\!000$\\\hline
\end{tabular}}\\\begin{tabular}{|l|l|l|}
\cline{2-3}\multicolumn{1}{c|}{$$}&\multicolumn{2}{l|}{{\bf Time ($10^{-3}$ s)}}\\\hline
\multicolumn{1}{|c|}{{\bf N}$^{\mathbf{3}}${\bf MHV Amplitudes}}& {\tt bcfw}&\multicolumn{1}{c|}{{\tt GGT}}\\\hline
$\mathcal{A}_{7}^{(5)}(-,+,-,+,-,-,-)$&$1.1 $&$2,\!500$\\
$\mathcal{A}_{8}^{(5)}(-,+,-,+,-,+,-,-)$&$3.5 $&$97,\!000$\\
$\mathcal{A}_{9}^{(5)}(-,+,-,+,-,+,-,+,-)$&$12 $&$1,\!100,\!000$\\
$\mathcal{A}_{10}^{(5)}(-,+,-,+,-,+,-,+,-,+)$&$35$&$14,\!000,\!000$\\
$\mathcal{A}_{11}^{(5)}(-,+,-,+,-,+,-,+,-,+,+)$&$97$&$?$\\
$\mathcal{A}_{12}^{(5)}(-,+,-,+,-,+,-,+,-,+,+,+)$&$210$&$?$\\\hline
\end{tabular}\end{center}\vspace{-1cm}
\end{table}

A good illustration of the relative simplicity afforded by momentum-twistor variables is the comparison between {\tt bcfw} and {\tt GGT} when using Drummond and Henn's recursive scheme, so that both packages are computing essentially the same functions. Indeed, merely translating the formulae for amplitudes given in \cite{Drummond:2008cr} into momentum twistors would seem to offer a remarkable improvement. This has in fact been done for NMHV \cite{ArkaniHamed:2009dn,Mason:2009qx} and N$^2$MHV \cite{Bullimore:2010pa} tree-amplitudes, but not more generally (although we suspect it would not be very difficult).

To see how merely  {\it translating } the spinor-helicity formulae written by Drummond and Henn could result in a remarkable increase inefficiency, observe that when written in terms of spinor-helicity variables and dual coordinates, each term in an N$^k$MHV amplitude will typically involve (see, e.g. \cite{Drummond:2008cr}) a large number of generalized spinor-helicity angle-brackets of the form `$\ab{a_1|x_{a_{1}a_{2}}x_{a_{2}a_{3}}\cdots x_{a_{k}a_{k+1}}|a_{k+1}}$', with $(k+1)$ region momenta sandwiched-between the spinors $\lambda_{a_{1}}$ and $\lambda_{a_{k+1}}$. This leads to an essentially boundless variety of new kinematical invariants that must be computed separately (the number of which grow very rapidly as $k$ increases). In contrast, as described in \mbox{section \ref{kinematics_section}}, the number of momentum-twistor four-brackets that can occur for tree-amplitudes is strictly bounded. To understand the magnitude of this handicap, consider that the expression used by {\tt GGT/S@M} to compute the 10-point N$^3$MHV alternating-helicity tree-amplitude involves more than 256-thousand {\it distinct} kinematical functions;\footnote{Here, we quote the number of distinct spinor-helicity invariants that would be used by the package {\tt S@M} for numerical evaluation.} in contrast, the formulae obtained with the Drummond and Henn recursion scheme in the momentum-twistor Grassmannian involves only 130 separate four-brackets.\footnote{Of course, these invariants are not all independent: they are related in complicated ways through momentum conservation. The fact that momentum-conservation is trivial for momentum-twistor variables has a dramatic impact on the simplicity of the formulae that result.} This makes the difference between 35 milliseconds and 4 hours given in \mbox{Table \ref{particular_evaluation_times}} much easier to comprehend. 


\section{Conclusions}

We have described a general, versatile, and efficient implementation of the tree-level BCFW recursion relations within the framework of {\sc Mathematica} which has been realized by the {\tt bcfw} package which is included with the submission  of this posting on the {\tt arXiv}.\footnote{From this note's abstract page on the {\tt arXiv}, choose the link to download ``other formats" (below the option for PDF) and you will find the {\tt bcfw} package and its associated walk-through file with many examples included in the `source' for this note. Also, you can download {\tt bcfw} at the project's page on {\tt http://hepforge.org}, where it will be generally maintained by the author.} 

Having access to an efficient, reliable, flexible, and robust toolbox for computing scattering amplitudes in $\mathcal{N}=4$ has proven an essential resource, and a important source of theoretical `data.' It is hard to overlook the exciting recent advances that have been made in our understanding of scattering amplitudes, and many of these results have relied heavily on being able to decisively rule-out or quickly confirm a wide-array of new ideas and proposals, leading to many new insights, and helping to establish what has a chance to become a fundamentally new descriptions of quantum field theory. 

We hope that the {\tt bcfw} package proves itself useful to a wide range of researchers---both as a reliable and efficient black-box for computing amplitudes, and as an educational resource for gaining intuition about the still somewhat unfamiliar, but extremely powerful new tools available to describe amplitude such as the momentum-twistor Grassmannian that have played an important role in the recent development of our understanding of scattering amplitudes in $\mathcal{N}=4$.\\

\noindent{\bf Acknowledgements}

We are extremely grateful for the opportunity to acknowledge the help of many people. In particular, we thank Jaroslav Trnka along with Henriette Elvang and Michael Kiermaier for their detailed help testing, debugging, and substantially improving {\tt bcfw}, and for many helpful discussions with Zvi Bern, Herald Ita, and Kemal Ozeren regarding the development of these tools. We thank Nima Arkani-Hamed, Freddy Cachazo, Simon Caron-Huot, and Matt Strassler for many helpful comments both on early drafts of this note, and regarding the interface and functionality of the {\tt bcfw} package.

\newpage
\appendix
\section{Glossary of Functions Defined by {\tt bcfw}}\label{glossary_of_functions}
\subsection{BCFW Recursion in the Grassmannian}
\begin{itemize}
\item {\tt bcfwPartitions[n,m]:} gives a list of the pairs $\{(n_L,m_L),(n_R,m_R)\}$ which should be bridged-together to compute the tree-level $n$-point N$^{(m-2)}$MHV amplitude $\mathcal{A}_n^{(m)}$ according to \[\mathcal{A}_n^{(m)}=\displaystyle\sum_{\substack{(n_L,m_L)\\(n_R,m_R)}}\mathcal{A}_{n_L}^{(m_L)}\underset{\mathrm{BCFW}}{\bigotimes}\mathcal{A}_{n_R}^{(m_R)}.\]

\item {\tt generalBCFWbridge[a,b,c][$\{\mathtt{n}_{\mathtt{L}},\mathtt{m}_{\mathtt{L}}\},\{\mathtt{n}_{\mathtt{R}},\mathtt{m}_{\mathtt{R}}\}$]:} computes the contributions to a given tree-amplitude arising from the term $\mathcal{A}_{n_L}^{(m_L)}\underset{\mathrm{BCFW}}{\bigotimes}\mathcal{A}_{n_R}^{(m_R)}$, where the arguments of $\mathcal{A}_{n_L}^{(m_L)},$ $\mathcal{A}_{n_R}^{(m_R)}$ and $\mathcal{A}_{n}^{(m)}$ have been `rotated' relative to $(1,\ldots,n_{\star})$ by $a,$ $b$ and $c$, respectively. There are at least two ways to envision how these different recursive schemes are defined. As illustrated in the figure on the left below, one can view the left- and right-side amplitudes as being actively rotated by amounts {\tt a} and {\tt b}, respectively---where rotation means that the legs which are deformed at the next stage of the recursion are offset relative to the default positions. On the right-hand figure, we view the same legs as being deformed at every stage, but at each stage of the recursion the amplitudes on the left- and on the right- have been cyclically-relabeled according to the figure (which induces the `rotations' of the first description). 
\vspace{-1cm}\begin{center}\raisebox{0.64cm}{\includegraphics[scale=0.5]{general_BCFW_2.pdf}}\includegraphics[scale=0.7]{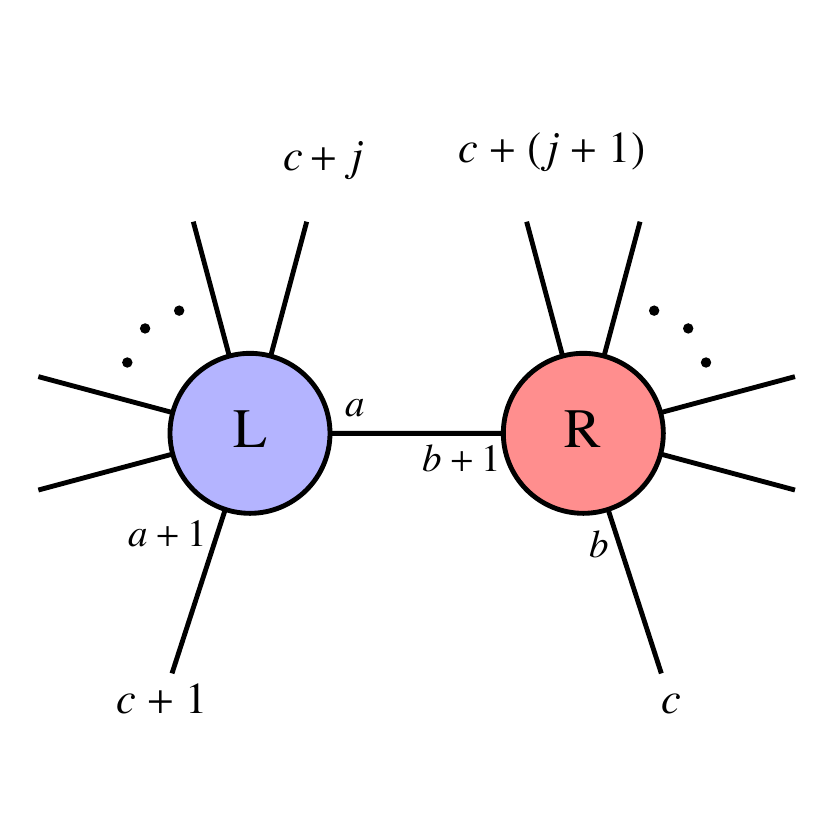}
\vspace{-0.4cm}\end{center}

\item {\tt generalTreeContour[a,b,c][n,m]:} the complete supersymmetric tree-amplitude $\mathcal{A}_n^{(m)}$---given in the representation obtained from the BCFW recursion scheme {\tt generalBCFWbridge[a,b,c]}. The contour is given as a list of contributing terms of the form \[\{\mathtt{residue},\mathtt{dMatrix}\},\] where each {\tt dMatrix} represents a point in the Grassmannian $G(k,n)$---in the gauge appropriate to the form of {\tt residue}. The full amplitude is given by the contour integral over the Grassmannian which `encloses' each of these poles; the contour integral's contribution from each pole is its corresponding {\tt residue}. A given helicity-component of each residue is obtained by multiplying the function {\tt residue} by the appropriate minors of its {\tt dMatrix}---which can also be viewed as the matrix of coefficients of the $\eta$'s appearing in the Fermionic $\delta$-functions.

\item {\tt treeContour[n,m]:} the default representation of supersymmetric tree-amplitudes. Specifically, $\mathtt{treeContour[n,m]=generalTreeContour[0,0,0][n,m]}.$
\end{itemize}

\subsection{Component Amplitude Extraction}
\begin{itemize}
\item {\tt Amp[helicityConfiguration]:} returns an analytic formula for the amplitude specified by {\tt helicityConfiguration}, obtained using the default BCFW recursion scheme. Specifically, \eq{\hspace{-1cm}\nonumber{\tt Amp[helicityConfiguration]=Total[AmpTerms[0,0,0][helicityConfiguration]].}}

\item {\tt AmpTerms[a\underline{~}:0,b\underline{~}:0,c\underline{~}:0][helicityConfiguration]:} generates each BCFW term for the specified {\tt helicityConfiguration} analytically, using the recursive scheme specified by {\tt a,b,c}. We should emphasize that formulae given by {\tt AmpTerms} are not particularly well-suited for numerical purposes: each term contains a great-deal of redundancy that would be systematically removed by the function {\tt nAmpTerms}.

\item {\tt mhvPrefactor[n]:} the universal pre-factor for all amplitudes obtained within the momentum-twistor Grassmannian integral: \eq{{\tt mhvPrefactor[n]:=}\frac{1}{{\tt ab[1,2]\cdots ab[n,1]}}.}

\item {\tt parseInput[helicityConfiguration]:} parses the input to the functions {\tt Amp}, {\tt AmpTerms}, {\tt nAmp}, and {\tt nAmpTerms}, which for an N$^{m-2}$MHV amplitude returns a list of four $m$-tuples which specify the four $(m\times m)$-minors of each {\tt cMatrix} which project-out the desired helicity configuration. Consider for example the amplitude \mbox{$\mathcal{A}_{8}^{(4)}\left(\psi_{+1/2}^{(1)},\psi_{+1/2}^{(1)},\psi_{+1/2}^{(1)},\phi_0^{(13)},\psi_{-1/2}^{(234)},\psi_{-1/2}^{(234)},\psi_{-1/2}^{(234)},\phi_0^{(24)}\right)$} of \mbox{Figure \ref{eight_point_example}}. That amplitude was computed via the command \eq{{\tt Amp\big[\{1\}, \{1\}, \{1\}, \{1, 3\}, \{2, 3, 4\}, \{2, 3, 4\}, \{2, 3, 4\}, \{2\, 4\}\big]};}
the arguments of the function {\tt Amp} were first parsed by {\tt parseInput}, returning 

\mathematica{0.95}{\hspace{-0.25cm}parseInput\big[\{1\},\{1\},\{1\},\{1, 3\},\{2,3,4\},\{2,3,4\},\{2,3,4\},\{2\,4\}\big]}{\hspace{-0.25cm}\{\{2,1,3,4\},\{5,6,7,8\},\{4,5,6,7\},\{5,6,7,8\}\}}

meaning that the specified helicity component-amplitude is obtained by multiplying each {\tt residue} in the {\tt treeContour} by \eqs{\hspace{-0cm}&\phantom{\,\times\,}{\tt Det\big[cMatrix_i[[All,\{2,1,3,4\}]]\big] Det\big[cMatrix_i[[All,\{5,6,7,8\}]]\big]}\\&\times {\tt Det\big[cMatrix_i[[All,\{4,5,6,7\}]]\big] Det\big[cMatrix_i[[All,\{5,6,7,8\}]]\big]}.}

If the input is ill-formed, apparently in error, or otherwise confusing (to {\tt bcfw}), {\tt parseInput} will print an error message and return `$-1$.'

\end{itemize}

\subsection{Input, Translation, and Random-Generation of Kinematical Data}\label{input_of_data}
\begin{itemize}

\item{\tt Qab[n]:} the matrix defined in equation (\ref{qab_definition}) which provides the map from momentum-twistors' components $\mu_a$ to the associated spinor-variables $\tilde\lambda_a$ via \eq{\tilde\lambda_a=\qab_{ab}\mu_b.}

\item{\tt QabInverse[n]:} the matrix defined in equation (\ref{qInverse_definition})---a ``formal inverse'' of $\qab_{ab}$---which provides a canonical map from spinor-helicity variables $\tilde\lambda_a$ to momentum-twistor components $\mu_a$. Of course, because $\qab_{ab}$ is a singular map, a particular representative of the one-to-many map $\qabinverse$ has was chosen in order to make equation (\ref{momentum_twistor_x_def}) literally correct once we have fixed the origin for the space of dual coordinate to be the point $x_1$.

\item {\tt setupUsingFourMomenta[inputFourMomentaList]:} translates a given list of momentum-conserving four-momenta (given as an $n$-tuple of four-vectors) into a standard set of spinor variables {\tt Ls}$\equiv\vec{\lambda}\equiv\left\{\lambda_a\right\}_{a=1\ldots n}$ and {\tt Lbs}$\equiv\vec{\tilde\lambda}\equiv\{\tilde{\lambda}_a\}_{a=1\ldots n}$ ---each, a globally-defined $(n\times 2)$ matrix---together with a set of corresponding momentum-twistors {\tt Zs}$\equiv\vec{\mathcal{Z}}\equiv\{\mathcal{Z}_a\}_{a=1\ldots n}$.

\item {\tt setupUsingSpinors[inputLs,inputLbs]:} translates a given list of momentum-conserving spinor variables {\tt inputLs} and {\tt inputLbs} (to become {\tt Ls} and {\tt Lbs}), each given as an $(n\times 2)$ matrix, into a standard set momentum-twistor variables {\tt Zs}, given as a globally-defined $(n\times 4)$-matrix.

\item {\tt setupUsingTwistors[inputTwistorList]:} translates a given list of generic momentum-twistors (given as an $(n\times 4)$ matrix) into a standard set of momentum-conserving spinor variables {\tt Ls}$\equiv\vec{\lambda}\equiv\left\{\lambda_a\right\}_{a=1\ldots n}$ and {\tt Lbs}$\equiv\vec{\tilde\lambda}\equiv\{\tilde{\lambda}_a\}_{a=1\ldots n}$ ---each, a globally-defined $(n\times 2)$ matrix. 

\item{\tt useRandomKinematics[n]:} randomly-generates on-shell, momentum-conserving four-momenta for $n$-particles in Minkowski signature, and sets up essentially all kinematical variables of potential interest, including 
\begin{itemize}
\item momentum-twistors $\{\overrightarrow{Z}\}\equiv${\tt Zs}, given as an $(n\times 4)$-matrix---the $n$ rows listing the four homogeneous components of each momentum-twistor; \linebreak {\tt useRandomKinematics[n]} also defines the `{\it dual}' momentum-twistors \linebreak $\{\overrightarrow{W}\}\equiv${\tt Ws}, which, although not used by {\tt bcfw}, may be found useful for some researchers;
\item spinor-helicity variables $\{\overrightarrow\lambda\}\equiv${\tt Ls} and $\{\overrightarrow{\tilde\lambda}\}\equiv${\tt Lbs}, each an $(n\times 2)$-matrix of components; these have been normalized so that $\tilde\lambda_a=\pm\left(\lambda_a\right)^*$, as described in \mbox{section \ref{kinematics_section}};
\item {\tt fourMomenta}, an $(n\times4)$ matrix of the components $(p^0,p^x,p^y,p^z)$ of each four-momentum;
\item {\tt regionMomenta}, the dual-coordinates (described in \mbox{section \ref{kinematics_section}}), given as a $n$-length list of $2\times2$ Hermitian matrices;
\end{itemize}

\item{\tt useRandomSpinors[n]:} the same random-kinematics-engine as that behind {\tt useRandomKinematics}, but much faster because it neglects to define any of the superfluous kinematical quantities (which {\it are} defined by {\tt useRandomKinematics}, making it rather inefficient).
\item{\tt useRandomTwistors[n]:} an extremely fast, minimalistic kinematics-generator that picks random, integer-valued twistors, and uses these to define the corresponding $\{\overrightarrow\lambda\}\equiv${\tt Ls}, etc. As with {\tt useReferences[n]}, the kinematical data generated using {\tt useRandomTwistors[n]} will neither be canonically normalized, nor will they correspond to {\it real} four-momenta in Minkowski signature.
\newpage
\item{\tt useReferences[n]:} uses a standard references-set of momentum-twistors to define external kinematics; these reference twistors were carefully selected so that \begin{itemize}
\item all components are integer-valued and relatively small;
\item there are no physical or spurious singularities;
\item all kinematical invariants are uniformly {\it positive} (that is, $s_{a\ldots b}>0$ for all ranges $a\ldots b$), and these invariants are numerically given by rational numbers composed of small integers---leading to amplitudes that are ratios of integers that are `not-too-horrendously-long.'
\end{itemize}
$\qquad$ The first few reference momentum-twistors are listed in \mbox{Table \ref{reference_momentum_twistors}}. Because an arbitrary set of momentum-twistors define on-shell, momentum-conserving kinematics, there are no constraints from momentum conservation. Therefore, choosing simply the fist $n$ from the list in \mbox{Table \ref{reference_momentum_twistors}} will suffice.

\end{itemize}
\subsection{Numerical Evaluation of Tree Amplitudes}
All of the functions below assume that the global variables such as the list {\tt Zs} have been appropriately defined by the user as described in section \ref{input_of_data}.
\begin{itemize}
\item {\tt nAmptTerms[a\underline{~}:0,b\underline{~}:0,c\underline{~}:0][helicityConfiguration]:} The numerically-optimized analogue of {\tt AmpTerms}. This function will return the numeric expressions which contribute to the given amplitude. It has built-within it several optimizations which (often dramatically) improve the evaluation of a given super-amplitude. Researchers interested in exporting the general computational strategy of the {\tt bcfw} to more efficient computational environments should try to mimic the functionality of {\tt nAmpTerms} rather than using the analytic formulae generated by {\tt AmpTerms}. 
\item {\tt nAmp[helicityConfiguration]:=Total[nAmpTerms[0,0,0][helicityConfiguration]}

\item{\tt toN:} includes the replacement rules necessary to numerically evaluate each momentum-twistor four-bracket and associated two-bracket. If the global list of momentum twistors {\tt Zs} have been defined, then {\tt (expression)/.toN} will convert {\tt expression} {\tt to} {\tt N}umbers.
\end{itemize}

\subsection{\AE sthetics}
\begin{itemize}

\item{\tt nice[expression]:} formats {\tt expression} by replacing {\tt ab[x$\cdots$y]}$\mapsto\ab{x\cdots y}$, and by writing any level-zero matrices in {\tt MatrixForm}. 

\item{\tt niceTime[timeInSeconds]:} converts a time measured in seconds {\tt timeInSeconds}, to human-readable form. 

\mathematica{0.75}{niceTime[299792458]\hspace{14cm}$~$
niceTime[10$^{{\tt-14}}$]}{{\tt 9.51 years}\hspace{14cm}$~$ {\tt 10.0 fs}}

\item{\tt order[expression,(option)]:} the the optional option {\tt option} set to {\tt 1}---its default value---{\tt order} will order all angle-brackets to a canonical ordering which prioritizes $x$-like pairing of arguments in four-brackets, and cyclically-ordered two-brackets (picking up any necessary signs from the necessary permutations). For example, 

\mathematica{0.8}{order[ab[1,3,4,9]]}{-ab[3,4,9,1]}

Notice that {\tt order} has {\it guessed} that the expression should be $\mathbb{Z}_9$-ordered (if there had been more brackets in {\tt expression} involving particle-labels larger than {\tt 9}, {\tt order} would have chosen the maximum argument of all {\tt ab} for cyclic ordering. 

If the optional argument {\tt option} were set to {\tt 0}, then {\tt order} will lexicographically-order the arguments of all angle-brackets, picking-up all necessary signs from the permutations involved.

\item{\tt twistorSimplify[expression,(option)]:} a general-mess-of-a function, which uses {\tt order} together with {\tt FullSimplify} endowed with the power of the most elementary, three-term Schouten identity \eq{\ab{X\,a\,b}\ab{X\,c\,d}+\ab{X\,b\,c}\ab{X\,a\,d}+\ab{X\,c\,a}\ab{X\,b\,d}=0} (where $X$ is any bi-twistor) to simplify expressions. {\tt twistorSimplify} can occasionally yield favourable results---especially when {\tt expression} is relatively simple, but---like {\tt FullSimplify}---is liable to exhaust patience. 

\item{\tt withTiming[expression]:} an all-purpose timing-wrapper, which will print-to-screen the time required to evaluate any expression using human-readable units of time (see {\tt niceTime}). {\tt withTiming} will also identify the function being evaluated in the output, and includes special formatting for the functions {\tt nAmp}, {\tt Amp}, {\tt nAmpTerms}, etc.

\end{itemize}

%

\providecommand{\href}[2]{#2}\begingroup\raggedright\endgroup

\end{document}